\documentclass[showpacs,preprintnumbers,amsmath,amssymb,twocolumn,nofootinbib]{revtex4}
\usepackage{graphicx}
\usepackage{dcolumn}
\usepackage{float}
\usepackage{bm}
\usepackage{hyperref}
\usepackage{epsfig,slashed}
\usepackage{verbatim}

\newcommand{\nco}{\newcommand}
\nco{\beq}{\begin{equation}} \nco{\eeq}{\end{equation}}
\nco{\beqa}{\begin{eqnarray}} \nco{\eeqa}{\end{eqnarray}}
\nco{\lra}{\leftrightarrow}
\def\sfrac#1#2{{\textstyle{#1\over #2}}}

\nco{\sss}{\scriptscriptstyle} \nco{\dphi}{\varphi}
\nco{\lsim}{\mbox{\raisebox{-.6ex}{~$\stackrel{<}{\sim}$~}}}
\nco{\gsim}{\mbox{\raisebox{-.6ex}{~$\stackrel{>}{\sim}$~}}}
\nco{\etal}{\textit{et al.}}
\nco{\ud}{\mathrm{d}}

\begin{document}


\title{Metastable dark matter mechanisms for INTEGRAL
511 keV $\gamma$ rays and DAMA/CoGeNT events}

\author{James M.\ Cline, Andrew R.\ Frey, Fang Chen}

\affiliation{%
\centerline{Physics Department, McGill University,
3600 University Street, Montr\'eal, Qu\'ebec, Canada H3A 2T8}
e-mail: fangchen, jcline, frey\ @physics.mcgill.ca }

\date{August 10, 2010}

\begin{abstract}  

We explore dark matter mechanisms that can simultaneously explain the
galactic 511 keV gamma rays observed by INTEGRAL/SPI,  the
DAMA/LIBRA  annual modulation, and the excess of low-recoil dark
matter candidates observed by CoGeNT.  It requires three nearly
degenerate states of dark matter in the 4-7 GeV mass range, with
splittings respectively of order an MeV and a few keV.   The top two
states have the small mass gap and transitions between them, either
exothermic or endothermic, can  account for direct detections. 
Decays from one of the top states to the ground state produce
low-energy positrons in the galaxy whose associated 511 keV gamma
rays are seen by INTEGRAL.   This decay can happen spontaneously, if
the excited state is metastable  (longer-lived than the age of the
universe), or it can be triggered by inelastic scattering of the
metastable states into the shorter-lived ones.  We focus on a
simple model where the DM is a triplet of an SU(2) hidden sector
gauge symmetry, broken at the scale of a few GeV, giving masses of
order $\lsim$ 1 GeV  to the dark gauge bosons, which mix kinetically
with the standard model hypercharge.   The purely decaying scenario
can give the observed angular dependence of the 511 keV signal 
with no positron diffusion, while the inelastic scattering mechanism
requires transport of the positrons 
over distances $\sim 1$ kpc before annihilating. We note
that an x-ray line of several keV in energy, due to single-photon
decays  involving  the top DM states, could provide an additional
component to the diffuse x-ray background. The model is testable by
proposed low-energy fixed target experiments.  

\end{abstract}

\pacs{98.80.Cq, 98.70.Rc, 95.35.+d, 12.60Cn}
\maketitle

\section{Introduction}

Annihilation of positrons near the galactic center gives rise to  a
narrow 511 keV gamma ray line that was first observed in 1972
\cite{Diehl}, and which has been confirmed by numerous experiments
since then, most recently by the SPI spectrometer aboard the INTEGRAL
satellite \cite{spi}.  The signal has two distinct components, one
associated with the central region of the galaxy (bulge) and  another
with the disk. There is as yet no consensus as to a conventional
astrophysical origin for these gamma rays \cite{astro,higdon,asym}, which evidently originate
from excess positrons annihilating nearly at rest. The apparent axial
symmetry of the bulge component is a point in favor for
proposals of models of dark matter (DM) that decays or annihilates into
low-energy positrons, since DM should be distributed symmetrically
near the galactic center.\footnote{In addition, the east-west
asymmetry in the disk component claimed by \cite{asym} is not confirmed
by the more recent analysis of \cite{bouchet2}, using more accumulated
data from INTEGRAL.}
  However early proposals of this sort were
driven toward DM candidates that were
nearly as light as the electron itself \cite{mevdm}, since the injection energy of
the positrons can be no greater than a few MeV \cite{Yuksel} (see
however \cite{chern}). 
Models of MeV scale dark matter that couples to $e^+e^-$ are highly
constrained  by low energy collider data, and are not (in our opinion)
theoretically attractive.  If dark matter is the source of 511 keV
gamma rays, one will need to verify its properties by direct
detection or other complementary means to make the explanation of the
511 keV signal convincing.

\begin{figure}[t]
\smallskip \centerline{\epsfxsize=0.5\textwidth\epsfbox{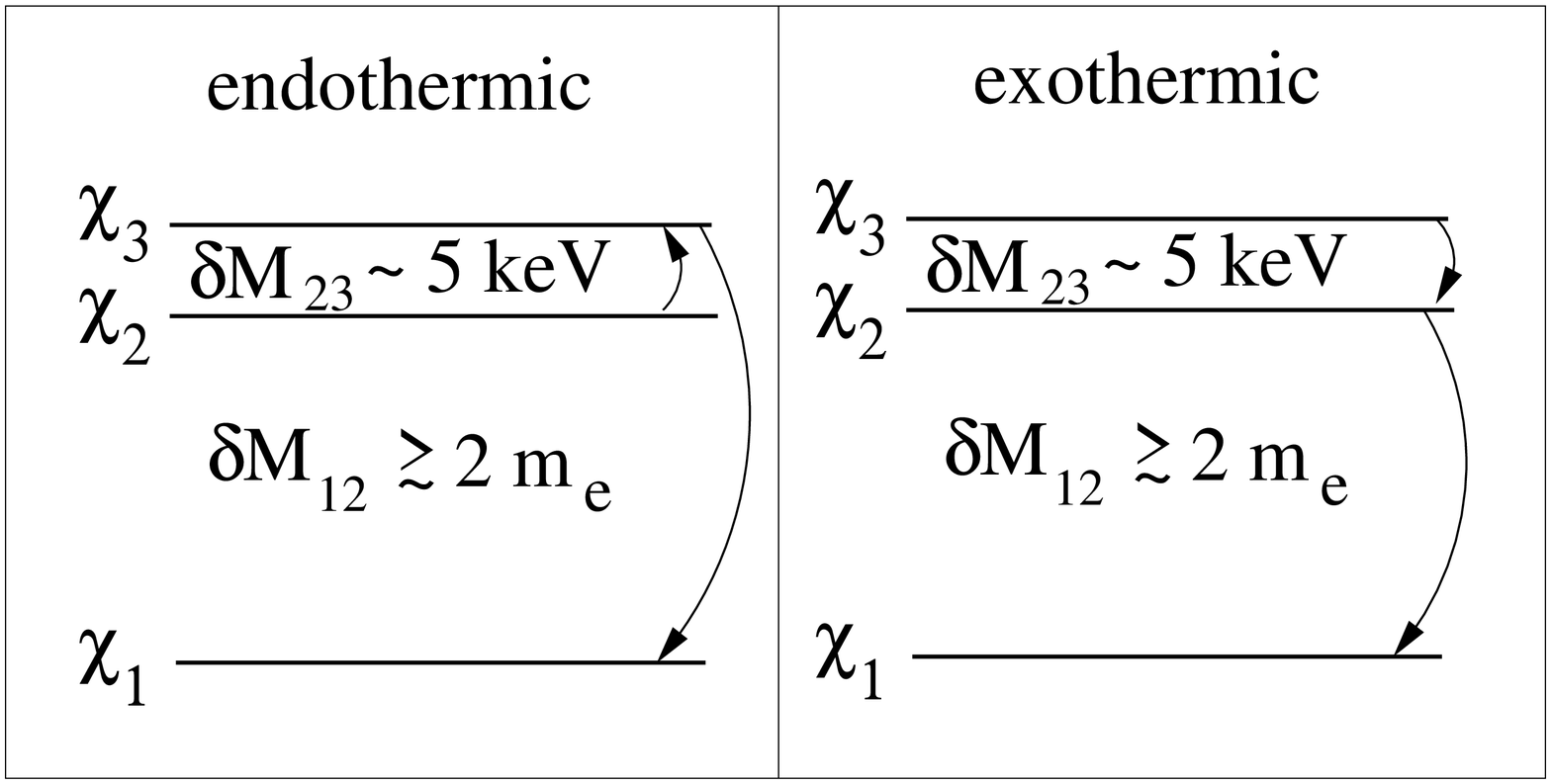}}
\caption{Spectrum of states for metastable dark matter models.  
Curves with arrows indicate the
sequence of transitions for the endothermic case, $\chi_2\to\chi_3\to
\chi_1$ (left) and the exothermic one, $\chi_3\to\chi_2\to\chi_1$
(right).}
\label{spect}
\end{figure}

In this work we propose and revisit scenarios in which a long-lived 
excited
state of DM with mass at the 10 GeV scale can scatter into a nearby
unstable state, whose mass differs by only a few keV.  The unstable
state decays into the ground state with the emission of a low-energy
$e^+ e^-$ pair.  The decay can be relatively fast, but the energy is 
only released after the slow process
of inelastic DM-DM collisions occurs.  (See however the purely
decaying variant described below.)  
There are two qualitatively different ways to realize
this, depending upon whether the metastable state is the  middle one,
requiring endothermic scattering,  or the top one, leading to
exothermic. The mass spectra and sequence of transitions  are
sketched in figure \ref{spect}.  The endothermic version,
in the context of 500-1000 GeV DM, was
first  presented in \cite{twist} to try to simultaneously explain
INTEGRAL, PAMELA (Payload for Antimatter Matter Exploration and
Light-nuclei Astrophysics) \cite{pamela} and 
ATIC (Advanced Thin Ionization Calorimeter) \cite{atic} excess electron
observations.  It was subsequently
discussed with applications to direct DM detection in 
\cite{Finkbeiner-metastable}, for $\sim 100$ keV values of the
small mass splitting.  The viability of the scenario for INTEGRAL
was further explored in \cite{us}, but only in the heavy DM
regime.\footnote{The original idea of excited dark matter assumed
that only the ground state was significantly populated, so that
excitation through the $\gsim$ MeV mass gap must occur in galactic
inelastic collisions \cite{xdm}.  However more detailed computations
showed that the collision rate is not high enough with such a large
energy barrier to overcome \cite{PR,twist,us}.}

It would clearly be interesting if the DM mechanism for the
INTEGRAL observations was somehow tied to direct detection of the DM
\cite{nima,Finkbeiner-metastable}.
Our exothermic mechanism is  partly motivated by 
ref.\ \cite{harnik}, which proposed a
model involving only the states $\chi_2$ and $\chi_3$ (in our
notation), as a means of
explaining two indications of direct detection of dark matter, namely
the long-standing DAMA/LIBRA annual modulation \cite{dama}, and the more recent
observation of excess low-recoil events by the CoGeNT 
(Coherent Germanium Neutrino Technology) experiment
\cite{cogent}. 
Ref.\ \cite{harnik} showed that DM with a mass of $\cong 4$ GeV and mass
splitting of a few keV  could be consistent with these observations,
using the exothermic nuclear scattering $\chi_3 N\to \chi_2 N'$. 
Their observation is that the shape of recoil spectrum is sensitive
to modulations of the local DM velocity when the scattering is
exothermic, and this can explain the DAMA observations.  Additionally
the overall rate for the same parameters is correct for explaining
the excess CoGeNT events.  The idea of ref.\ \cite{harnik} is related
to the inelastic dark matter proposal \cite{idm}, which however
assumed the scatterings to be endothermic rather than exothermic.
(See \cite{Batell-multicomponent} for another discussion of exothermic scatterings.)

There have been several proposals for DM in the 5-10 GeV mass range
to explain the DAMA and CoGeNT observations \cite{lightdm}.   
Most recently, ref.\ \cite{hooper} showed that elastic DM interactions
could simultaneously explain the DAMA/LIBRA and CoGeNT observations
if the dark matter mass is near 7 GeV and the cross section on
nucleons around $2\times 10^{-40}$ cm$^2$.  We will argue that our
endothermic scenario is close enough to being elastic, if the small
mass splitting $\delta M_{23}$ is of order a few keV, so that the
same analysis applies.  Even though such a small splitting has little
effect on DM-nucleus scattering, it is important for DM-DM scattering
in the galaxy, in the present case where the DM is lighter than the
nuclei in the direct detectors.  Getting the observed rate of galactic
positrons limits the maximum mass splitting in this case, to values
somewhat lower than those that would strongly affect the direct detection
rates.

\begin{figure*}[t]
\smallskip \centerline{\epsfxsize=\textwidth\epsfbox{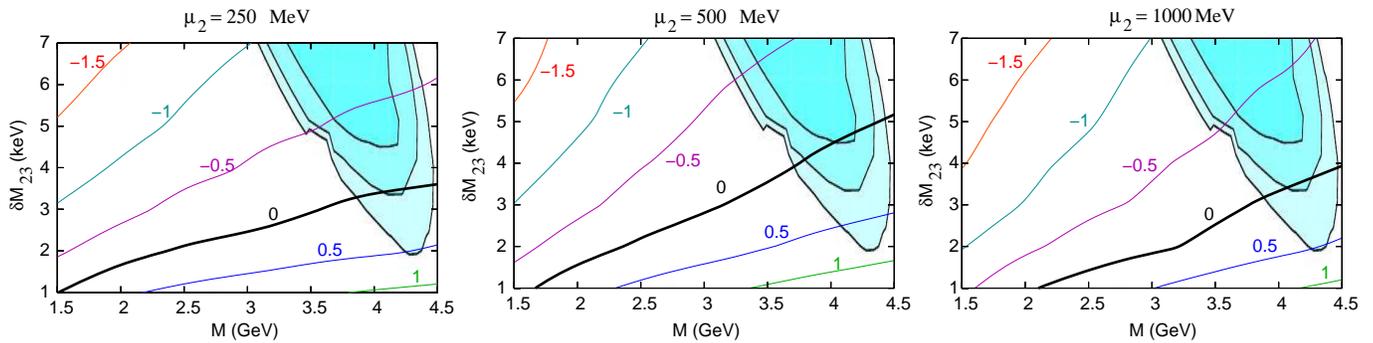}}
\caption{Solid curves: contours of
$\log R_{e^+}/R_{\rm obs}$  (the positron production rate)
for exothermic dark matter, in the 
plane of the average DM mass $M_\chi$ and mass splitting $\delta M_{23}$.
 Heavy contour labeled ``0'' matches the observations.  Shaded regions
are allowed by DAMA/LIBRA, from ref.\ \cite{harnik}.  Each plot takes
a different value of  gauge boson mass $\mu_2$, with $\mu_{1,3}$ given
by eqs.\ (\ref{mu1eq},\ref{mu3eq}).  DM halo parameters are given
by eq.\ (\ref{einasto-params}).}
\label{exothermic}
\end{figure*}

An interesting variant of the above mechanisms is to assume that the 
unstable excited state is so long-lived that it still has a relic 
population in the galaxy, and so does not need to be produced by DM
collisions.\footnote{Earlier work on decaying DM as the source of
511 keV gamma rays can be found in \cite{decays,PR}.}\ \   This version has more freedom, in that the rate of 
producing positrons  (via decays into $\chi_1 e^+ e^-$) does not
depend upon the small mass splitting $\delta M_{23}$, whereas the
rate of inelastic scattering 
$\chi_3\chi_3\leftrightarrow\chi_2\chi_2$ is rather sensitive to
$\delta M_{23}$.  

Our proposals fit nicely within the framework of dark matter with a
nonabelian gauge symmetry in a hidden sector, as  suggested by
\cite{nima}, since such DM automatically consists of multiple states,
and small mass splittings are radiatively generated if the gauge
symmetry is spontaneously broken.   The simplest example that
contains three DM states is a hidden SU(2) gauge sector, where the DM
is in the triplet representation.  After the hidden SU(2) breaks, 
two colors of the dark  gauge
boson must acquire small kinetic mixing  $\epsilon_{i}\sim
10^{-3}-10^{-6}$ with the standard model hypercharge $Y$,
\beq
		{\cal L}_{\rm mix} = \sum_{i}\epsilon_i B_i^{\mu\nu}
	Y_{\mu\nu}
\label{kinmix}
\eeq
while the remaining one must
have negligible mixing to keep the long-lived state stable against
decays to $\chi_1 e^+ e^-$.  The mixing $B_{i}$'s couple weakly
to charged Standard Model (SM) particles, 
and mediate the scatterings with nucleons or
decays into $e^+e^-$.\footnote{Except in equation (\ref{lag}), we will
rescale the $\epsilon_i$ to be the mixing parameter of $B_i$ with the
photon for notational convenience.} 
We find that the hidden gauge symmetry should break at
the 10 GeV scale (resulting in gauge boson masses of order several
hundred MeV) to give the right cross sections for DM scattering in
the galactic center and in detectors.  An attractive feature of these
hidden sector gauge boson masses and couplings is that they are in
the right range to be directly probed by new proposed fixed-target
experiments \cite{BPR}.

We will present our main results first, in section \ref{results}.
The remaining parts of the paper supply the many details leading
to these results.  Section \ref{models} specifies the hidden sector
SU(2) particle physics models we consider.   
The gauge coupling $\alpha_g$ of this SU(2) is calculated in section
\ref{relic_density} by the requirement of getting the right thermal
relic density for the DM.  There we also work out the crucial 
relative abundances of the excited states.  Section \ref{rate-angle}
describes how the rate and angular distribution of 511 keV gamma rays
are computed.  Here we also summarize what is believed about the
location of gaseous media in the galactic bulge where positron annihilation is
supposed to take place, in respose to criticisms of DM interpretations of the
INTEGRAL observations in ref.\ \cite{higdon}.
In section \ref{ddrates} we explain how the gauge
kinetic mixing parameter $\epsilon_1$ is constrained to get the 
desired rates for DAMA.  Various astrophysical constraints are 
addressed in section \ref{constraints}.  Our predictions for
the masses and couplings of the gauge boson $B_1$ that mediates
the DAMA and CoGeNT reactions, relevant for direct laboratory
searches, are presented in section \ref{lab}.  We conclude in 
\ref{conc}.  The appendices give further details about the kinetic
equilibrium of the DM with the SM, and the cross section for DM 
annihilation.

\section{Main results} 
\label{results}

In this section we summarize our main results. The details leading up to
them will be given in subsequent sections. The relevant parameter space is
the average DM mass $M_\chi$, the mass splitting $\delta M_{23}$, and the
masses $\mu_{i}$ of the hidden sector gauge bosons $B_{i}$ that mediate the
interactions with the standard model.   We fix the larger mass splitting to
be $\delta M_{12} = 1.1$ MeV so that there is sufficient phase space for
the decay into $\chi_1 e^+ e^-$ while insuring that the decay products are
not very relativistic, as required by constraints on the injection energy
of the low-energy positrons \cite{Yuksel}.  Larger values of $\delta
M_{12}$ tend to  suppress the positron rate, and the direct detection rate
for exothermic DM, because of greater depletion of the  excited state
abundance, but our  results are not greatly sensitive to the exact  value
so long as it is less than a few MeV.  

Because the  
nonabelian SU(2) gauge interactions take the form
\beq
 g\bar \chi_1 \slashed B_2 \chi_3 + {\rm cyclic\ permutations,}
\eeq
$B_1$ mediates the transition $\chi_3\to\chi_2$, {\it et cyc.}
The strength of the gauge coupling $g$ is fixed by the requirement of
getting the observed relic density of DM from thermal
freeze-out,
\beq
	\alpha_g \cong c_g \left(1 - {\bar\mu^2\over M_\chi^2}\right)^{-1/4}{M_\chi\over {\rm GeV}}
\label{relic_alpha}
\eeq
where $\alpha_g = g^2/4\pi$ and $\bar\mu$ is the average mass of the
gauge bosons.
The value of the constant $c_{g}$ depends upon the number of 
hidden sector Higgs
bosons that can be present in the final state of $\chi\chi \to HH$
annihilations;  it can lie in the range $c_g\cong (1.7-2.5)\times 10^{-5}$ 
for the scenarios we consider.    In the following, 
we assume the dark Higgs bosons are
heavier than the DM, which yields the top value in this range,
hence a larger rate of positron
production.  The value of $c_g$  is derived  in section 
\ref{relic_density}.  

\begin{figure*}[t]
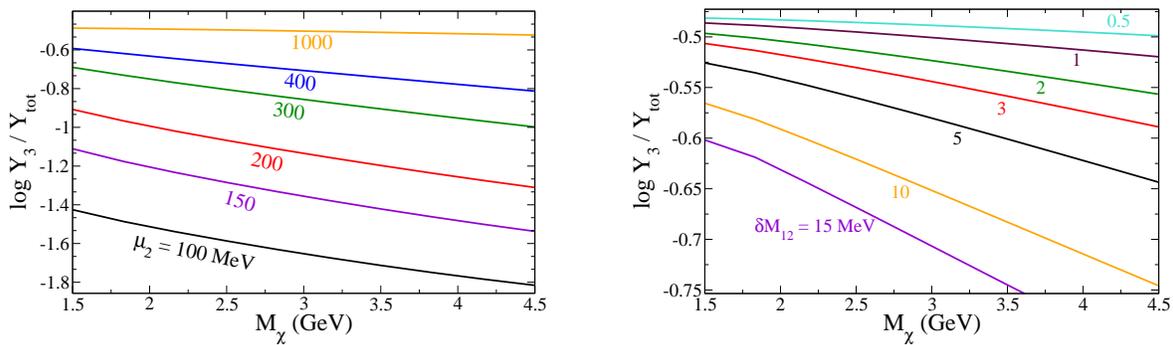

\smallskip \centerline{
\epsfxsize=0.4\textwidth\epsfbox{abundance4.eps}\hfil
\epsfxsize=0.4\textwidth\epsfbox{abundance-dM3.eps}
}
\caption{Left: in the exothermic case, log of $Y_3/Y_{\rm tot}$, 
abundance of  stable excited state 
$\chi_3$ relative to
the total DM abundance, as a function of $M_\chi$, for
several values of the gauge boson mass $\mu_2$, with $\mu_{1,3}$ 
fixed as in eq.\ (\ref{mu1eq},\ref{mu3eq}) and $\delta M_{23} = 5$ keV.  $\delta M_{12}$
is fixed at 1.1 MeV.  Right:  same but with varying $\delta
M_{12}=0.5-15$ MeV, and fixed $\mu_2=1000$ MeV.}
\label{abundance}
\end{figure*}

\begin{figure}[t]
\smallskip \centerline{
\epsfxsize=0.4\textwidth\epsfbox{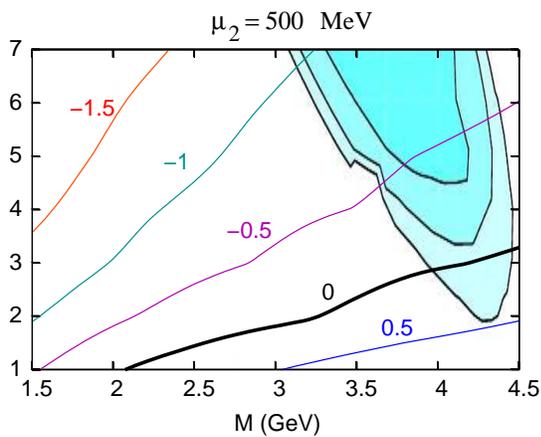}}
\caption{As in fig.\ \ref{exothermic}, but for less cuspy Einasto
profile with $\alpha = 0.08$, $r_s=7.5$ kpc, $\rho_\odot = 0.42$
GeV/cm$^3$.}
\label{nocusp-down}
\end{figure}

\begin{figure*}[t]
\smallskip \centerline{\epsfxsize=\textwidth\epsfbox{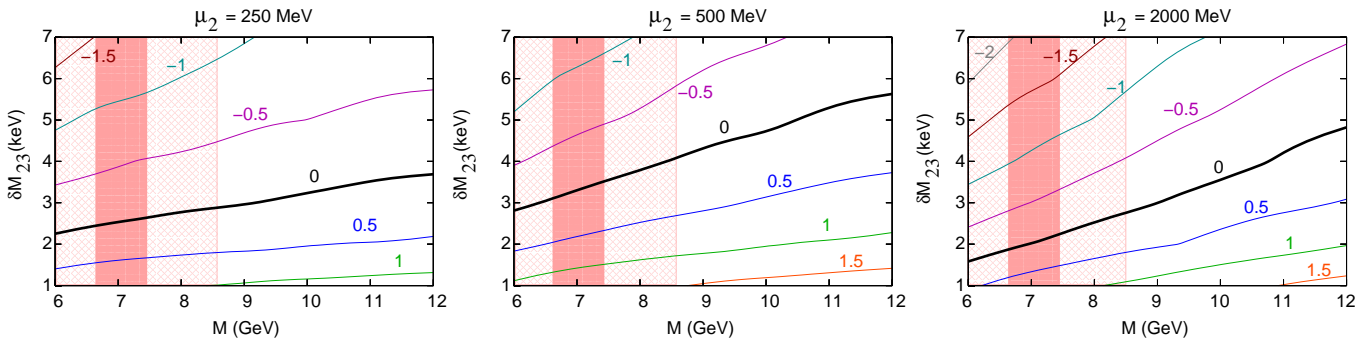}}
\caption{Solid curves: contours of
$\log R_{e^+}/R_{\rm obs}$  (the positron production rate) 
for endothermic dark matter, analogous to fig.\ \ref{exothermic} for the exothermic
case.  Einasto halo parameters are given in (\ref{einasto-params2}).
Columns correspond to gauge
boson masses $\mu_2 = 250$, 500 and 1000 MeV respectively.  Shaded
regions are 90\% and 99\% c.l.\ preferred DM masses for fitting DAMA/CoGeNT data, from
ref.\ \cite{hooper}.}
\label{endothermic}
\end{figure*}

\begin{figure}[t]
\smallskip \centerline{
\epsfxsize=0.4\textwidth\epsfbox{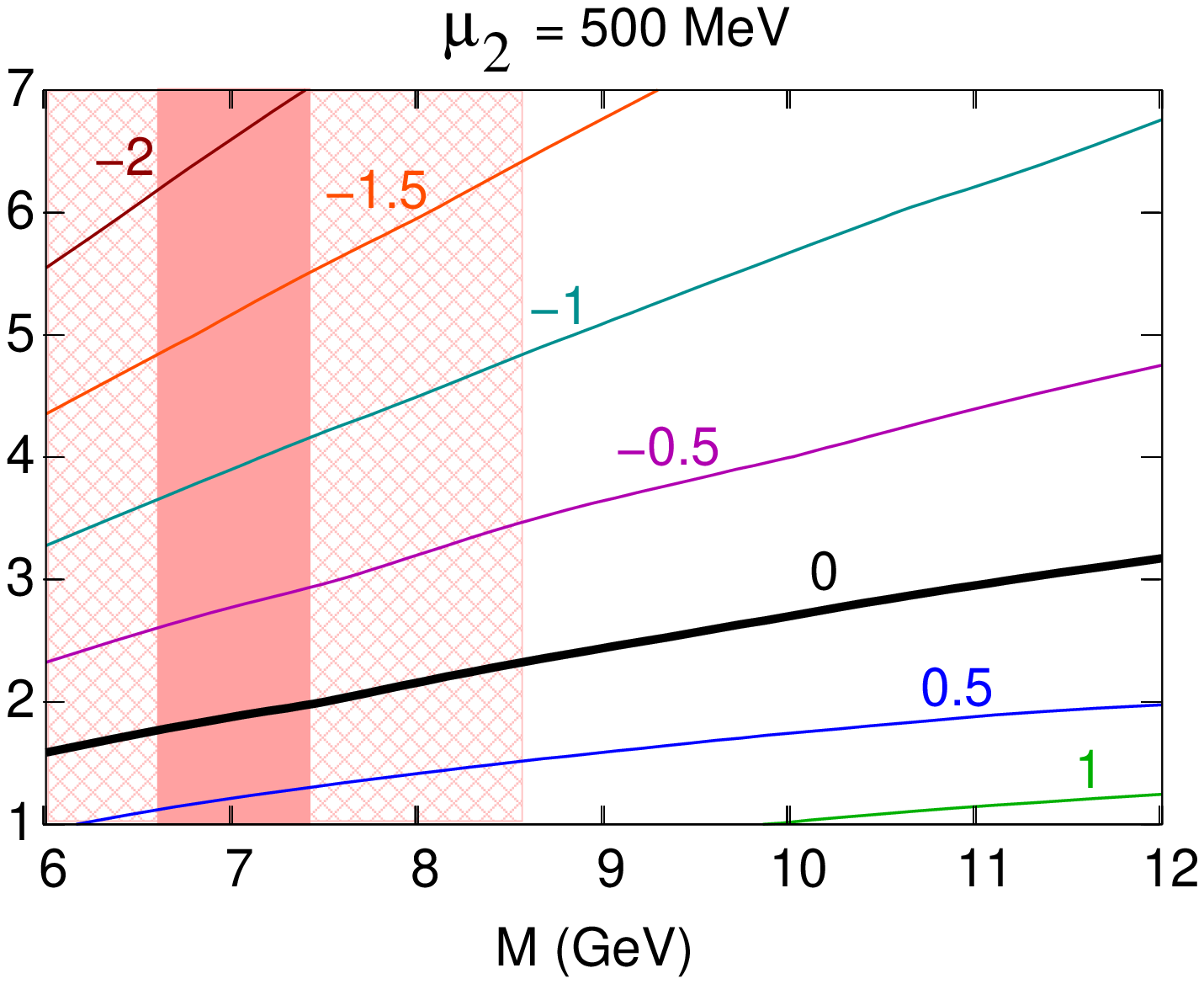}}
\caption{As in fig.\ \ref{endothermic}, but for less cuspy Einasto
profile with $\alpha = 0.12$, $r_s=12$ kpc, $\rho_\odot = 0.42$
GeV/cm$^3$.}
\label{nocusp-up}
\end{figure}

\begin{figure*}[t]
\smallskip \centerline{
\epsfxsize=0.4\textwidth\epsfbox{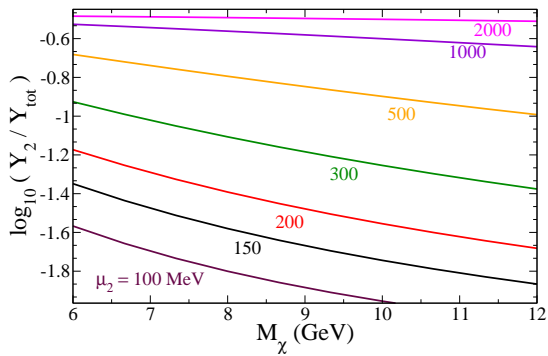}\hfil
\epsfxsize=0.4\textwidth\epsfbox{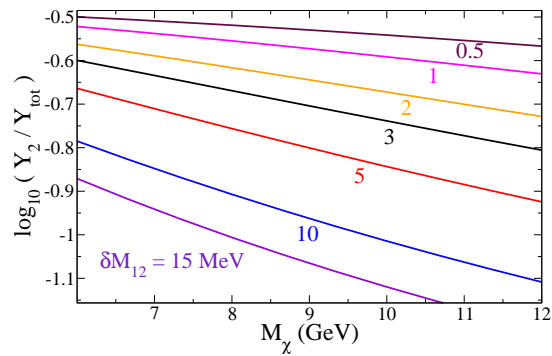}
}
\caption{Similar to figure \ref{abundance}, but for endothermic
model.}
\label{abundance-up}
\end{figure*}

\subsection{Exothermic dark matter}

We first consider the exothermic case where $\chi_3$ is the stable
excited state.
In figure \ref{exothermic} we plot contours of $\log R_{e^+}/R_{\rm obs}$,
the predicted rate of positron production at the galactic center
versus the measured rate, in the $M_\chi$-$\delta M_{23}$ plane,
where $\delta M_{23}$ is the small mass splitting between DM states
2 and 3.  
The contours are superimposed upon the DAMA/LIBRA allowed region of ref.\ 
\cite{harnik}. To illustrate the dependence on the gauge boson  
masses, each graph has a different value of $\mu_2$, the mass of 
$B_2$.  In the exothermic case, $B_2$ is the color that has negigible mixing
with the SM (to avoid $\chi_3\to\chi_1$ decays), 
and so $\mu_2$ does not directly affect the rates of
either direct detection nor galactic positron production.  However,
the class of models we describe in section \ref{models} predicts
relations between $\mu_2$ and the other gauge boson masses,
\beqa
\label{mu1eq}
	\mu_1 &\gsim & \sqrt{\mu_3^2-\mu_2^2}\\
	\mu_3 &=& {2\over\alpha_g}\delta M_{23} + \mu_2
\label{mu3eq}
\eeqa
The first condition (\ref{mu1eq}) depends on details of how the Higgs 
mechanism in the hidden sector gives masses to the gauge bosons;
we take the inequality to be saturated, which helps to increase the
rate of $\chi_3\chi_3\to\chi_2\chi_2$ scatterings (since $B_1$ is
the exchanged boson).  
The second condition (\ref{mu3eq}) arises because the mass difference
$\mu_3-\mu_2$ determines the radiatively generated splitting $\delta
M_{23}= -\frac12\alpha_g(\mu_3-\mu_2)$.   

From figure \ref{exothermic} one observes that larger values of
$\mu_2$ help to achieve a large enough rate of positron production,
up to some optimal value $\mu_2\sim 600$ MeV, beyond which the rate
starts to slowly fall with $\mu_2$. The rise at small $\mu_2$ occurs 
because increasing $\mu_2$  inhibits $\chi_3\chi_3\to\chi_1\chi_1$
downscatterings in the early universe, lessening the depletion of the
$\chi_3$ state. Figure \ref{abundance}, left panel, illustrates this
more directly, where the relic abundance of the excited state $Y_3$
relative to that of total DM abundance  $Y_{\rm tot}$ is plotted as a
function of $M_\chi$ for several values of $\mu_2$.  There is a
saturation $Y_3/Y_{\rm tot}\to\ \sim 1/3$ as $\mu_2$ approaches the GeV scale,  for
the fiducial value $\delta M_{12} = 1.1$ MeV of the large mass
splitting.   (The right panel of fig.\  \ref{abundance} indicates that
this saturation would occur at higher values of $\mu_2$  if $\delta
M_{12}$ is increased.  The rate of $\chi_3\chi_3\to\chi_1\chi_1$
increases with $\delta M_{12}$ due to the  larger phase space.)
Further increase of $\mu_2$ beyond the optimal point decreases
the positron rate, because $\mu_1$ is an increasing function of 
$\mu_2$, and the rate of $\chi_3\chi_3\to\chi_2\chi_2$ transitions
goes like $\mu_1^{-4}$.

It may seem surprising that the rate of positron production  is a
decreasing function of the mass splitting $\delta M_{23}$, since the
phase space for $\chi_3\chi_3\to\chi_2\chi_2$ increases with $\delta
M_{23}$.  However, so does the exchanged momentum that appears in the
propagator of the virtual gauge boson, and this has the more
important effect of suppressing the amplitude; see eq.\
(\ref{sigma_down}).

Figure  \ref{exothermic} shows some overlap between the desired rate
of positron production and the DAMA allowed region for $\mu_2 \gsim 200$
MeV. For each point in the $M_\chi$-$\delta M_{23}$ plane, we have
adjusted the value of $\epsilon_1$ to obtain the DAMA dectection rate
assumed by ref.\ \cite{harnik}.  $\epsilon_3$ is taken
to be $\lsim 10^{-3}$; the results shown are insensitive to the
exact value.  Concerning $\epsilon_3$, 
an intriguing prediction of our
model is that each direct detection of the process $\chi_3 N\to
\chi_2 N'$  must be accompanied by the subsequent production of
$e^+e^-$ through the decay $\chi_2\to\chi_1 e^+ e^-$ (whose rate
scales as $\epsilon_3^2$), so in principle
one could look for the positron in coincidence.  However, the
lifetime for the decay cannot be much less than $10^3$ s, as we will
show in section \ref{unstable}.   Since the speed of DM in the galaxy
is of order $10^{-3}c$, this occurs too far from the experiment to
detect the $e^+e^-$ pair.  In fact this lifetime is much longer
than the age of the universe for $\epsilon_3 \lsim 10^{-8}$, leading
to an alternative possibility for explaining the 511 keV signal via
decays of primordial $\chi_2$, more about which in section \ref{ddmsc}. 

The rate of positron production through DM excitations is sensitive to the density
profile $\rho$ of the DM halo; it scales like $\rho^2$ evaluated
near the galactic center.  We parametrize the shape using the Einasto profile  
\beq
\rho= \rho_\odot  \exp\left(-\frac2{\alpha}\left((r/r_s)^\alpha 
-(r_\odot/r_s)^\alpha\right)\right) 
\label{einasto}
\eeq 
A set of values that are often
considered to be standard are $\alpha=0.17$, $r_s=20$ kpc, $\rho_\odot =
0.3$ GeV/cm$^3$, $r_\odot = 8.5$ kpc.  These values for $\alpha$ and $r_s$ are based upon pure
dark matter $N$-body simulations that do not not take into account
the effects of baryons in the central region of the galaxy
\cite{Navarro}.
We do not obtain a large enough
rate of positron production using these numbers.  However, there is
strong evidence that the halo is much more concentrated (cuspy) near the
center  than these values indicate, due to the presence of
the baryons, which have the effect of contracting the density
\cite{bdm}.
Table \ref{tab1} shows the profile parameters for six Milky Way-like
galaxies from the Aquarius simulation, which have been reanalyzed
to include baryonic contraction \cite{tissera}.  
Furthermore it has been argued that the local density may  be
larger than the canonical value by a  factor of $1.3-2$
\cite{pato,salucci}.  We find that
the exothermic scenario gives acceptable overlap between the INTEGRAL
and DAMA-allowed regions only if we
adopt a DM halo that is very cuspy and has a somewhat large density in the solar 
neighborhood.  We take the most concentrated example in table
\ref{tab1},
\beq
	\alpha = 0.065,\ r_s = 5.3 {\rm\ kpc},\ \rho_\odot = 0.42{\rm\
GeV/cm}^3
\label{einasto-params}
\eeq
to obtain fig.\ \ref{exothermic}.  This could
still be considered a conservative choice, 
since ref.\ \cite{salucci} argues for
$\rho_\odot = 0.43(11)(10){\rm\ GeV/cm}^3$.  With these error
estimates, one might reasonably consider 
$\rho_\odot = 0.6 {\rm\ GeV/cm}^3$.  This allows for 
some reduction of the cuspiness of the
halo with very similar results, to $\alpha=0.08$, $r_s = 7.5$ kpc
for example. 
Moreover, we can still achieve reasonable consistency using the
same 
cuspy profile 
while keeping $\rho_\odot = 0.42$ GeV/cm$^3$; see fig.\
\ref{nocusp-down}.

\begin{table}[b]
  \begin{center}
  \caption{Characteristics of the density 
    profiles of the haloes in the
    Aquarius galaxy formation simulations of ref.\ \cite{tissera}. 
    Column 1 gives the name
    of each halo.  Columns 2-3 list $\alpha$, $r_{s}$, 
the parameters of the best fitting Einasto
    model, in the inner region of the galaxy.}
\label{tab1}
\begin{tabular}{|l|c|c|}\hline
\ {Galaxy} \ & \ { $\alpha$}\  &\  { $r_{s}$ (kpc)}\   \\\hline
Aq-A-5 &  0.065 &  \phantom{1}5.3 \\	
Aq-B-5 &  0.145 &  15.6\\
Aq-C-5 &  0.115&  10.2  \\
Aq-D-5 &  0.102 & 14.7 \\
Aq-E-5 &  0.098 &  11.1 \\
Aq-F-5 &  0.112 &  15.6\\
\hline
\end{tabular}
 \end{center}
\vspace{1mm}
\end{table}

\begin{figure}[t]
\smallskip \leftline{
\epsfxsize=0.4\textwidth\epsfbox{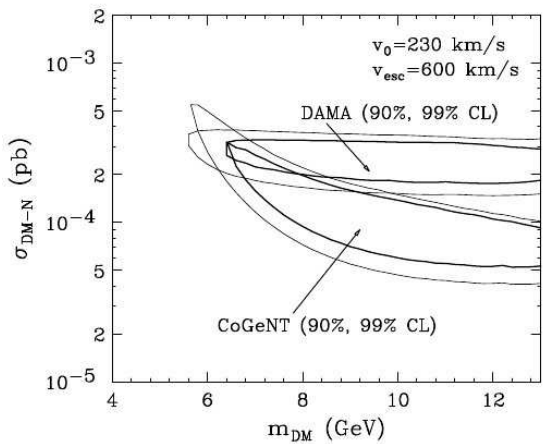}}
\caption{Allowed regions of ref.\ \cite{hooper} for DM to explain
DAMA/LIBRA and CoGeNT events.}
\label{hooper_fig}
\end{figure}

\subsection{Endothermic dark matter}

If $\chi_2$ is the stable state, then the transitions $\chi_2\to\chi_3$ are
endothermic. The energy barrier would tend to reduce the rate of such
transitions compared to the exothermic case, but there are other
differences that also affect the rate.  Most importantly, even though eqs.\
(\ref{mu1eq},\ref{mu3eq}) are unchanged, the roles of the gauge bosons
$B_2$ and $B_3$ become interchanged relative to exothermic DM. $\mu_3$ now
controls the rate of $\chi_2\chi_2\to\chi_1\chi_1$ downscattering in the
early universe, hence the relic density of $\chi_2$. Because $\mu_3$ is
naturally the heaviest of the three gauge boson masses in our model (see
section \ref{models}), this means that the endothermic scenario
leads to a significantly larger rate of galactic positrons than the
corresponding exothermic one.  We thus adopt a less cuspy halo profile
in this case,
\ref{tab1},
\beq
	\alpha = 0.08,\ r_s = 8 {\rm\ kpc},\ \rho_\odot = 0.42{\rm\
GeV/cm}^3
\label{einasto-params2}
\eeq
Our findings for the 511 keV signal for endothermic DM
are illustrated in figure \ref{endothermic},
We have the freedom to choose even less cuspy
profiles if desired, with some accompanying decrease in the value
of $\delta M_{23}$, as shown in fig.\ \ref{nocusp-up}, using the 
more moderate parameter values $\alpha=0.12$ and $r_s=12$ kpc.
Fig.\ \ref{abundance-up} shows how the
relative abundance of the stable state, $Y_2/Y_{\rm tot}$, 
depends upon the
masses $M_\chi$, $\mu_2$ and mass splitting $\delta M_{12}$.
For the examples shown, $\delta M_{23}$ should
be $\lsim 4$ keV to match the direct detection rate corresponding
to fig.\  \ref{hooper_fig},
in the allowed $M_\chi$ region that is shaded in figs.\ 
\ref{endothermic}, \ref{nocusp-up}.

Similarly to the exothermic case, we fix the value of $\epsilon_1$
to get the desired direct detection rate, while $\epsilon_2$, which
controls the rate of decay $\chi_3\to\chi_1 e^+ e^-$, is a free
parameter.  We assumed $\epsilon_2=10^{-3}$ in fig.\
\ref{endothermic}.  In contrast to the exothermic case, the results
are somewhat sensitive to this choice: taking much smaller values
of $\epsilon_2$ mildly suppresses the rate due to its effect on 
the relic abundance $Y_2$, as we will further discuss in section
\ref{kineq}.  

Ideally, the analysis of ref.\ \cite{hooper} should
be redone for our slightly inelastic scattering to see how the overlap
of the DAMA and CoGeNT allowed regions of fig.\ \ref{hooper_fig} might 
be modified.  (For this reason we display our results in the same
range of DM masses as in fig.\ \ref{hooper_fig}.)  We leave such an
investigation to future work.

\begin{figure}[t]
\smallskip \centerline{
\epsfxsize=0.45\textwidth\epsfbox{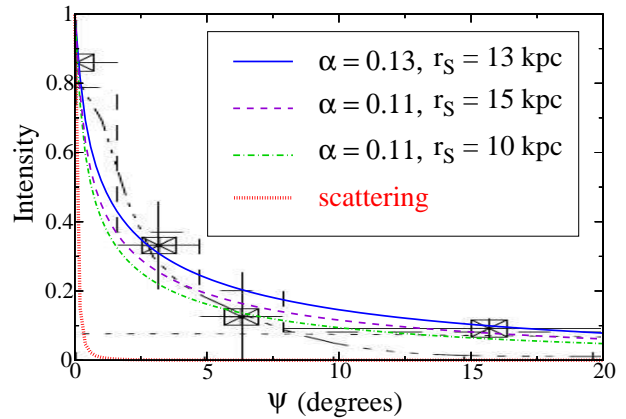}}
\caption{Data points show angular distribution from galactic center of observed 511 keV
signal, reproduced from ref.\ \cite{bouchet2}; labeled curves are 
predictions from DM scattering (lowest, dotted curve) and from decay (upper
curves), assuming no propagation of positrons before they annihilate.}
\label{angular}
\end{figure}

\subsection{Decaying DM scenario}
\label{ddmsc}

If the gauge mixing parameter $\epsilon_{2}$ or $\epsilon_{3}$ is sufficiently
small, then the excited state $\chi_{2}$ or $\chi_{3}$ 
(depending upon whether the DM is exothermic or
endothermic) which we have referred to
as  ``unstable''   can be as long lived as
the universe.   Let us denote the stable and ``unstable'' excited
states by  $\chi_s$ and $\chi_u$.   Instead of being produced in
$\chi_s\chi_s\to \chi_u\chi_u$ scattering, $\chi_u$ can have a
significant relic density and produce  $e^+ e^-$ from its slow
decays.   In section \ref{ddmrate} we show that the correct lifetime for
producing the observed rate of positrons results from taking
\beq
	\epsilon_{2,3}\sim 10^{-11}
\eeq
The exact expression depends upon other parameters and is given
by eq.\ (\ref{eps23eq}).  In particular, the dependence upon the
DM halo profile is much weaker for decays than for the inelastic
scattering scenarios discussed above.  We are no longer constrained
to consider profiles such as (\ref{einasto-params}).  

It is intriguing that for reasonable choices of the halo
profile, the decaying DM scenario can explain the morphology of the
511 keV signal without the need to invoke propagation of the
positrons before annihilation.  This is in contrast to the inelastic
scattering mechanisms which localize the positron production much
closer to the galactic center, as we next discuss.

\subsection{Angular profile of 511 keV signal}
\label{res-angular}

If positrons annihilate before propagating, 
the predicted intensity $I_{e^+}$ of the 511 keV signal as a 
function of 
angle is just a reflection of the DM density profile,\footnote{In the
case of inelastic scattering, there is some mild $r$-dependence
of $\langle \sigma v\rangle$ due to the $r$-dependence of the
velocity dispersion, which we neglect here.}\  through
a line-of-sight integral, whose form depends upon whether the
positrons were created through scattering or decay:
\beq
	I_{e^+}(\hat x) = 
	 \int_{\rm l.o.s.} dx \,\left\{ \begin{array}{ll}
        \frac12\langle \sigma v\rangle\, 
	{\rho_s^2\over 	M_\chi^2},& {\rm scattering}\\
	{\rho_u\over 	M_\chi\, \tau_s}, & {\rm decay}
	\end{array} \right.
\label{los}
\eeq
The integral is 	 
taken along the $\hat x$ direction, where $\rho_{s,u}$ is the density of 
the stable or unstable excited DM component $\chi_{s,u}$, proportional to the total density 
$\rho$, and $\tau_u$ is the lifetime of $\chi_u$.

For the Einasto profile (\ref{einasto-params})  we considered for
scatterings, $\rho^2$ is practically a delta function, and so the
signal would appear to come from a point source.   $\rho^2$ falls to
$e^{-8}$ of its maximum value at a distance of  $r_8 = r_s
(2\alpha)^{1/\alpha}$.  Even for the standard profile with
$\alpha=0.17$ and $r_s=20$ kpc, $r_8$ is only 35 pc, which subtends
an angle of $0.2^\circ$.  However the observed signal subtends at
least $8^\circ$ \cite{bouchet2}; see fig.\ \ref{angular}.  Therefore if the
scattering explanation is correct, all of the observed width must be
due to propagation.\footnote{Our results differ somewhat from those
of ref.\ \cite{Abidin}, which assumed a less cuspy halo.}\\  The distance corresponding to $8^\circ$ is 1.1
kpc, which may be astrophysically reasonable,  depending upon the 
structure of the galactic magnetic field and the injection energy of
the positrons.  

Apart from astrophysical mechanisms of positron transport
\cite{Prantzos,Jean}, which strongly depend upon the poorly constrained magnetic field
of the inner galaxy, our model
suggests another way in which this widening could occur due to the
streaming of $\chi_3$ ($\chi_2$ in exothermic case)  before it
decays.  If the gauge mixing parameter for $B_{2,3}$, the hidden
gauge boson mediating the $\chi_{3,2} \to\chi_1 e^+ e^-$ decay, is
sufficiently small, $\epsilon_{2,3}\sim 10^{-7}$, then $\chi_{3,2}$ 
is so long-lived that it will travel approximately 1 kpc before
decaying.  

For the decaying DM scenario, it is possible to fit the observed
angular distribution without any smoothing from positron diffusion.
Figure \ref{angular} shows several examples with Einasto parameters
$\alpha = 0.11$ and $\alpha = 0.13$ that pass through all the error bars. 
These examples are close to the ones given in table \ref{tab1}, and 
so could be considered realistic in light of baryonic compression of
the inner part of the DM halo.

\section{Particle physics models}
\label{models}
The simplest example of a nonabelian
hidden sector model consistent with the observations we discuss 
has a dark SU(2) gauge group under which the DM transforms as a
triplet.  The most general form of the Lagrangian that we will need is
\beqa
\label{lag} 
 {\cal L} &=& \sfrac12\bar\chi_a (i\slashed{D}_{ab}
-M_\chi\delta_{ab})\chi_b 
-\sfrac{1}{4 g^2} B^a_{\mu\nu} B_a^{\mu\nu} \\
&-& \sum_i\sfrac{1}{\Lambda_i}\Delta_a^{(i)}
B_a^{\mu\nu}Y_{\mu\nu} - \sfrac12 y\bar\chi_a \Sigma_{ab}\chi_b
+{\cal L}_{\rm Higgs}(\Delta^{(i)},\Sigma)\nonumber
\eeqa
Here $\Delta^{(i)}$ and $\Sigma$ are triplets and a quintuplet
respectively of the hidden SU(2).  Two such triplets
 are needed in order to get the 
kinetic mixing (\ref{kinmix}) required for the direct detection signal
and galactic positron production.  The mixing parameters $\epsilon_i =
\langle \Delta_i\rangle/\Lambda_i$ arise when the triplets acquire
VEVs.  These mixing parameters lead to a coupling $e \epsilon_i\cos\theta_W$
of the electric current to $B_i$; in the remainder of this paper, we
rescale $\epsilon_i$ to remove the Weinberg angle from this coupling.
A third triplet is required to get the right pattern of 
mass splittings for exothermic DM.  The VEV of the quintuplet 
$\Sigma_{ab}$ gives the large $\sim$ MeV mass splitting.

We studied this class of models previously in ref.\ 
\cite{nonabelian}.  It is convenient to take the triplet VEVs to be
mutually orthogonal $\langle \Delta_a^{(i)}\rangle \equiv \delta_{ia}
\Delta_a$.  
Ref.\ \cite{nonabelian} shows that it is easy to construct
a potential leading to this pattern.  We further restrict the
traceless symmetric tensor $\Sigma_{ab}$ to have VEVs only on the 
diagonal,
\beq
	\langle \Sigma \rangle = {\rm
diag}(A-B,\ 2B,\ -A-B),
\eeq
This alignment can be accomplished by suitable small interactions
between $\Sigma$ and the triplets.  With these VEVs, the masses of the
gauge bosons are given by
\beqa
	\mu_1^2 &=& g^2(\Delta_2^2 + \Delta_3^2 + 2(A+3B)^2)\nonumber\\
	\mu_2^2 &=& g^2(\Delta_1^2 + \Delta_3^2 + 8A^2)\nonumber\\
	\mu_3^2 &=& g^2(\Delta_1^2 + \Delta_2^2 + 2(A-3B)^2)
\eeqa
The corresponding mass shifts in the $\chi_{a}$ states relative to the
average mass $M_\chi$ are given by
\beqa
	\delta M_1 &=& -\frac{1}{2}\alpha_g\left(\mu_2+\mu_3\right)
        + y(A-B)
	\nonumber\\
	\delta M_2 &=& -\frac{1}{2}\alpha_g\left(\mu_1+\mu_3\right)
 	+ y(2B)
	\nonumber\\
	\delta M_3 &=& -\frac{1}{2}\alpha_g\left(\mu_1+\mu_2\right)
	+ y(-A-B)
\label{dMs}
\eeqa

We have introduced the Yukawa coupling contribution in order to explain two
different scales of mass splittings: $\delta M_{23}\sim $keV, and $\Delta
M_{12}\sim\delta M_{13}\sim $MeV.  This can occur if the quintuplet VEVs
satisfy  $A=-3B$; then the Yukawa term only contributes to the large mass
splittings and not to $\delta M_{23}$.   Let us assume this to be the case;
we will presently show how it can come about.  Then the small mass
splitting comes entirely from the one-loop self-energy contribution from
gauge boson exchange, $\delta M_{23} =  \delta M_3-\delta M_2 =
\frac12\alpha_g(\mu_3 - \mu_2)$. The assumed order $\delta M_3 > \delta
M_2$ requires that $\mu_3>\mu_2$, hence $\Delta_2 > \Delta_3$.  The most
economical choice would be to remove $\Delta_3$ from the spectrum
altogether, which is permissible if the kinetic mixing
parameter $\epsilon_3$ is allowed to vanish.  In fact for the endothermic
scenario, this is exactly what we want, in order to forbid 
$\chi_2\to\chi_1 e^+ e^-$ decays, so that $\chi_2$ can be 
stable.\footnote{See section \ref{life_metastable}, though, for further
discussion of some subtleties.}  For the exothermic case, it is
opposite: we need to insure the stability of $\chi_3$ against decays to
$\chi_1$,  hence $\epsilon_2$ must be negligible, while $\epsilon_3$ is
needed for the $\chi_2\to\chi_1 e^+ e^-$ decays. And for both scenarios,
$\epsilon_1$ must be nonzero to enable direct detection via inelastic
$\chi_{2,3}$ scattering on nucleons.  The upshot is that we need all three
triplets for exothermic DM (although only two of them should lead to
kinetic mixing), but only two, $\Delta_{1,2}$, for endothermic.

Now we return to the question of why the quintuplet VEVs should satisfy 
the seemingly fine-tuned relation $A=-3B$.
Interestingly, the desired VEVs can arise
from the simple renormalizable potential
\beq
	V(\Sigma) = \lambda_{\sss\Sigma} ({\rm tr}\Sigma^2 - v^2)^2
	+ \mu\det\Sigma
\eeq
which has three degenerate minima at $A=\pm 3B$ and $A=0$. (In the 
absence of the triplet VEVs, this  would leave one of the three gauge
bosons massless, breaking SU(2)$\to$U(1)).  We assume that it is
possible to design small
interactions with the triplets that align
$\langle\Sigma\rangle$ along the diagonal, and which might perturb 
$A$ slightly away from $-3B$.  For simplicity we take $A=-3B$ in the 
following, so that $\delta M_{12}\cong 6 y B$.  

Considering the smaller mass splitting, if $M_\chi\sim 5$ GeV, the
gauge coupling is of order $\alpha\sim 10^{-4}$, and the difference  
in gauge boson masses should be of order $|\delta M_{23}|/\alpha_g
\sim 100$ MeV.   This is consistent with triplet VEVs at the scale of
$\sim 30$ GeV, since the gauge coupling is $g\sim 0.035$.  To get the
correct sign for the mass difference, $M_3>M_2$ only requires that
$\Delta_2>\Delta_3$ (given our assumption $A=-3B$), so that
$\mu_3>\mu_2$.

\section{Relic density}
\label{relic_density}

In ref.\ \cite{nonabelian} a first attempt was made to compute the
value of the gauge coupling $\alpha_g$ corresponding to the observed
DM density through thermal freeze-out.  In this section we correct
and refine that result, taking into account some subleading effects,
including extra annihilation channels into dark Higgs bosons,
dependence of the Born cross section on the DM velocity, and
Sommerfeld enhancement \cite{sommerfeld} at the time of freeze-out.   Moreover we
estimate the amount of dilution of the excited states  due to
downscattering in the early universe. 

\subsection{Annihilation cross section; determination of $\alpha_g$}
\label{cross_alphag}

Ref.\ \cite{nonabelian} derived the annihilation cross section for
$\chi\chi\to B B$ by separately considering  $\chi_1\chi_1\to B_j
B_j$ (with $j=2,3$) and $\chi_1\chi_{j}\to B_1 B_{j}$ (again
with $j=2,3$), and explicitly
averaging over the initial state  and summing over the final state
colors.  In this paper, we make several improvements to the previous
calculation as well as minor corrections.  Details of our calculation
are given in appendix \ref{thermalrelic}.

First, we now include the process
where  $\chi\chi$ goes to two hidden sector Higgs bosons through exchange
of a virtual $B$  in the $s$-channel.  Furthermore, as pointed out in
\cite{feng}, velocity-dependence (including Sommerfeld enhancement) may
make significant contributions to the cross section in some models.  
Therefore, we include corrections to order $v^2$ in the tree-level 
cross section as well as the leading contribution from Sommerfeld 
enhancement.  

Moreover, in the present application, 
the dark matter is sufficiently light that
its annihilation may take place in the broken phase of the
theory.  Therefore, we have computed the cross section for $\chi\chi\to BB$
taking into account the gauge boson masses.  The main effect comes
simply from the reduction in phase space, which is a factor of
$(1-\mu_i^2/M_\chi^2)^{1/2}$ in the cross section, where $\mu_i$ is the
mass of the gauge boson in the final state.  Annihilation to light Higgs bosons
is also modified only by this factor (of course, Higgs bosons heavier than
the DM are not annihilation products). 
For simplicity we replace $\mu_i$ by the average mass $\bar\mu$ 
of the gauge bosons and light Higgs bosons in this part of the calculation.

The full cross section
for annihilation into gauge bosons, $N_3$ Higgs triplets, and $N_5$ Higgs
quintuplets, including leading velocity dependence, is
\beqa
	\sigma v_{\rm rel} &=& {\pi\over 12} {\alpha_g^2\over M_\chi^2}
	\left[\left(\frac{25}{2}+2N_3+10N_5\right)\left(1+
{\pi\alpha_g\over v_{\rm rel}}\right)\right.\nonumber\\
&+&\!\!\left.\left(\frac{317}{48} -\frac{5N_3}{12}-\frac{25N_5}{12}\right) 
v_{\rm rel}^2\right]
\left(1-{\bar\mu^2\over M_\chi^2}\right)^{1/2}
\label{sigv}
\eeqa
in the center of momentum frame.  
The factor $(1+{\pi\alpha_g/ v_{\rm rel}})$ incorporates Sommerfeld
enhancement neglecting the masses of the gauge bosons.  This neglect
is valid for large $M_\chi$, such that the freezeout temperature
$\sim M_\chi/20$ is above the symmetry breaking scale.  For smaller
$M_\chi$, this factor is roughly an upper bound on Sommerfeld 
enhancement (except very close to a resonance) and furthermore
$\alpha_g$ is sufficiently small that Sommerfeld
enhancement is an unimportant correction during freezeout.  Conversion to 
the rest frame of the cosmic fluid introduces an additional correction
of order $V^2$ for center of momentum velocity $\vec V$.

To compute the relic density, one needs the thermal average 
$\langle \sigma v_{\rm rel}\rangle$.  Using the Maxwell-Boltzmann
distribution, we find
\beqa
\langle \sigma v_{\rm rel}\rangle &=&  {\pi\over 12} 
{\alpha_g^2\over M_\chi^2}
\left[\left(\frac{25}{2}+2N_3+10N_5\right)\times\right.\nonumber\\
&&
\left. \left(1+\alpha_g \sqrt{\frac{\pi M_\chi}{T}}-\frac{1}{2\pi}
\frac{T}{M_\chi}\right)\right.
+ \\
&&\left. \left(\frac{317}{8} -\frac{5N_3}{2}-\frac{25N_5}{2}\right)
{T\over M_\chi}\right]
\left(1-{\bar\mu^2\over M_\chi^2}\right)^{1/2}\ .\nonumber 
\label{sigvavg}
\eeqa
Annihilations go out of equilibrium at a temperature given by
$M_\chi/T_f = x_f \cong \ln\xi -\frac12\ln\ln\xi$ with $\xi = 1.0\times
10^{12}(M/$TeV) for triplet DM, giving $x_f = 20.4$ for 
$M_\chi = 5$ GeV and $x_f = 23.3$ for $M_\chi = 100$ GeV. We should
equate (\ref{sigvavg}) at the $T_f$ with the cross section needed to
match current constraints on the DM density.  This varies mildly with
$M_\chi$ as $\langle\sigma v\rangle_0 \cong (3.2 - 0.24\,
\log(M/$GeV))$\times 10^{-26}$\,cm$^3$/s \cite{slatyer-com}.  However, the latter expression
assumes the usual particle content of the standard model at the time 
of freezeout, whereas in our model there are three additional gauge
bosons and extra dark Higgs bosons.  This increases both the 
Hubble rate and redshifting between freezeout and the present. Thereby,
the extra particle content decreases $\langle\sigma v\rangle_0$ by
a factor of $\sqrt{1 + (6+3N_3+5N_5)/61.75}$.  Details are given in appendix
\ref{thermalrelic}.  For two Higgs triplets and one quintuplet,
we find that a good approximation is given by
\beq
	\alpha_g \cong {1.7\times 10^{-5}\over
\left(1-{\bar\mu^2/ M_\chi^2}\right)^{1/4}}\,
 {\left(M_\chi/{\rm GeV}\right)\over
	\sqrt{1 + 7.7\alpha_g}}
\label{relic_ag}
\eeq
For the range of $M_\chi$ we are interested in, $\alpha_g$ is so small that
the Sommerfeld enhancement factor $\sqrt{1 + 7.7\alpha_g}$ can be neglected.
Further results for other light Higgs states are given in \ref{alphagfinal}.

\subsection{Relative density of excited state}

Let us denote the stable excited state by $\chi_s$, and the unstable one as
$\chi_u$.   At the freezeout temperature, all three DM states are equally
populated, but if the rate of downscattering transitions $\chi_s\chi_s\to
\chi_1\chi_1$ remains larger than the Hubble rate at temperatures below the
mass splitting $\delta M_{1s}$, the density $n_s$ of the excited state gets
suppressed relative to $n_1$ of the ground state.  The rate of
$\chi_s\chi_s\to\chi_u\chi_u$ transitions in the galaxy at the present
epoch scales with $(n_s/n_{tot})^2$ (for $n_{tot}=n_1+n_s+n_u$).  
We must therefore compute this ratio to
accurately predict the  rate of positron production.  The direct detection
rate for $\chi_s N\to \chi_u N'$ similarly scales like $(n_s/n_{tot})$.  

To compute the dilution of $\chi_s$ from downscattering, we solve the
Boltzmann equation for the abundance $Y_s = n_s/s$, where $s$ is the entropy
density.  Defining $z= \delta M_{1s}/T$, it can be cast in the form
\cite{semiann}
\beq
	{dY_s\over dz}  = -{\lambda\over z^2} \left(Y_s^2 - Y_1^2
\left(Y_s^{\rm eq}\over Y_1^{\rm eq}\right)^2\right)
\label{boltz1}
\eeq
where $\lambda$ is related to the cross section $\sigma_{\downarrow}$ for 
$\chi_s\chi_s\to\chi_1\chi_1$ downscattering
by
\beq
	\lambda = \langle \sigma_{\downarrow} v\rangle \left.{s\over
H}\right|_{z=1}
\eeq
except (as we shall describe below) the multiplicity factors $g_*$ and
$g_{*s}$ that appear in $\lambda$ should retain their $z$-dependence
(only explicit factors of $T$ get replaced by $\delta M_{1s}$).  

We can simplify this by assuming that the abundance of the ground state
does not change significantly during the depletion of $\chi_s$, so
$Y_1$ is just a constant.  Furthermore $Y_2^{\rm eq}/Y_1^{\rm eq} \cong
e^{-z}$ to a good approximation if the DM is in kinetic equilibrium with 
the standard model particles (we will discuss this caveat below).
Defining the fraction $f = Y_s/Y_1$,
(\ref{boltz1}) becomes
\beq
	{df\over dz} = -{\bar\lambda \over z^2}\left(f^2 -e^{-2z}\right)
\label{boltz2}
\eeq
where $\bar\lambda = \lambda Y_1$. (However the $e^{-2z}$ will be
modified when we take into account kinetic decoupling effects; see
next subsection.)   To explicitly compute $\lambda Y_1$,
let us parametrize the DM ground state density as $n_1 = 
(g_{*s}/g_{*s,0})\xi T^3$, where the $g_{*s}$ factors take into account the
dilution of $n_1$ as a function of temperature due to entropy production
after freezeout.   Then
\beq
	\bar\lambda = {g_{*s}\, \xi\, \delta M_{1s}\, M_p
	\over 1.66\sqrt{g_*}\,
	g_{*s,0}}\langle \sigma_{\downarrow} v\rangle
\eeq
where $M_p = 1.22\times 10^{19}$ GeV and $\xi = 7\times 10^{-10}$ GeV/$M_\chi$
to match the observed DM density.

The cross section $\sigma_{\downarrow}$ is straightforward to compute, since
it is similar to $e^- e^-$ scattering, with just two diagrams, exchange of
a gauge boson in the $t$ and $u$ channels.  In the low-velocity limit,
we obtain
\beq
	\langle\sigma_{\downarrow} v\rangle = 4\pi\alpha_g^2 {M_\chi^2 v_t
\over (\mu_i^2 + M_\chi^2 v_t^2)^2}
\label{sigma_down}
\eeq
where $v_t = \sqrt{2\delta M_{1s}/M_\chi}$ is the velocity of $\chi_1$
at threshold (when the incoming $\chi_s$ particles are at rest) and
$\mu_i$ is the mass of the exchanged gauge boson.  For the exothermic
DM model, where $\chi_s = \chi_3$, $i=2$, while for the endothermic case
where $\chi_s = \chi_2$, $i=3$.  Using this constant cross section is actually
a conservative estimate, as it is near the maximum value of the full
velocity dependent cross section given in equation (\ref{sigma_down_full})
for $\delta M_{1s}\sim 1.1$ MeV and our typical values of $M_\chi$, $\mu_i$.

To solve the Boltzmann equation, we first tried to 
employ the semianalytic technique
popularized in Kolb and Turner \cite{KT}.  Namely, one writes $f = e^{-z}
+\Delta$ and linearizes the equation in $\Delta$ for the early 
time behavior, giving $\Delta =
z^2/(2\bar\lambda)$, while $\Delta' = -\bar\lambda
z^{-2}\Delta^2$ at late times.  Integrating the latter equation between the
$z$ of freezeout, $z_f$, and infinity gives the final abundance
${Y_s\over Y_1} = \Delta_\infty = z_f/\bar\lambda$.
The trick then is to appropriately
determine the value of $z_f$.  One does this by assuming that 
\beq
\Delta(z_f) = {z_f^2\over 2\bar\lambda} = c e^{-z_f}
\label{zfeq}
\eeq
and then finding the value of $c$ for which this procedure
best reproduces the full numerical solution.  However we find that 
this procedure is not sufficiently accurate for the regime we are
interested in, where ${Y_s/Y_1}\gsim 0.1$ rather than the 
exponentially small values of interest for $Y_1$ itself.  There is
no fixed value of $c$ that accurately gives the same as the numerical
result as ${Y_s/Y_1}$ ranges between $0.1$ and $1$.  Therefore we numerically
solve the Boltzmann equation in all cases.

To relate $Y_s/Y_1$ to the fractional abundance of the stable state
to the total dark matter population at the present time, we must
remember that the unstable state  $\chi_u$  is also kept in
equilibrium with $\chi_1$ until a similar freezeout temperature
(which is the same in the limit that $\mu_2=\mu_3$).  
Only at much later times $>10^3$ s, $\chi_u$ decays to the
ground state.  The total abundance of dark matter is then 
$Y_1 + Y_u + Y_s$.  The fractional abundance of
$\chi_s$ is therefore
\beq
	{Y_s\over Y_{\rm tot}} \cong {Y_s/Y_1 \over 1 + Y_s/Y_1 + 
	Y_u/Y_1}
\label{Ys}
\eeq
where $Y_{s,u}/Y_1$ denotes the values at freezeout, from solving the
Boltzmann equation.  We compute $Y_{u}/Y_1$ in exactly the same
way as $Y_{s}/Y_1$.  The only difference is the exchange of
$\mu_2\leftrightarrow\mu_3$ for the gauge boson mass appearing in the
propagator of the cross section (\ref{sigma_down}).  An astute reader may
wonder whether $\chi_3\chi_3\to\chi_2\chi_2$ scatterings change the 
ratio $Y_u/Y_s$ additionally; however, the cross section (\ref{sigma_down})
is greatly reduced for $\delta M_{23}\sim 1-10$ keV, and we find like
\cite{harnik} that this process freezes out at temperatures well above
the mass splitting $\delta M_{23}$.

\begin{figure}[t]
\smallskip \centerline{\epsfxsize=0.2\textwidth\epsfbox{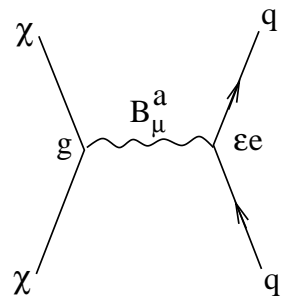}}
\caption{Scattering of $\chi$ on charged particle $q$ that 
keeps DM in kinetic equilibrium.} \label{scatt}
\end{figure}

\subsection{Kinetic equilibrium}
\label{kineq}

The preceding discussion of the Boltzmann equation assumed that the DM is in
kinetic equilibrium until the freezeout of downscattering.  If this is not
the case, the relic density of $\chi_s$ will be smaller than estimated
there.  The reason is that the equilibrium density depends upon the kinetic
temperature $T_k$ and this redshifts with the expansion of the universe 
$T_k\sim
1/a^2$, in contrast to the temperature of particles that are still coupled
to the thermal bath, $T\sim 1/a$.  To get some sense of the size of
the effect, we can follow the analytic procedure for an approximate
result, even though in the end we solve the Boltzmann equation
numerically. 

If $T_d$ is the kinetic
decoupling temperature, then $T_k = T^2/T_d$ for $T< T_d$.  Let $z_d =
\delta M_{1s}/T_d$.  Then the term $e^{-2z}$ in (\ref{boltz1}) must be
replaced by $\exp(-2\,{\rm max}(z, z^2/z_d))$.   Following the semianalytic
approach described above, one finds that eq.\ (\ref{zfeq}) is replaced by
\beq
	\Delta(z_f) = {z_f^3\over c\lambda z_d} = e^{-z_f^2/z_d}
\eeq
which can be rewritten as 
$z_f = (z_d \ln(c\bar\lambda z_d/z_f^3))^{1/2}$.  As a consequence
the relic abundance of 
$\chi_s$ is suppressed by $\sqrt{z_d}$ in this case.  Thus
it is preferable for kinetic decoupling to occur after the 
chemical freezeout of $\chi_s$, for maximizing its relic density.

\begin{figure}[t]
\smallskip \centerline{\epsfxsize=0.45\textwidth\epsfbox{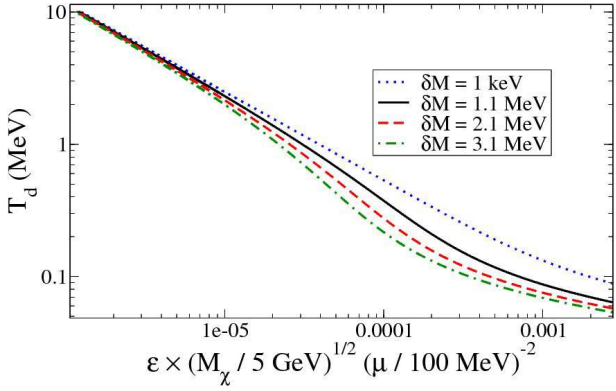}}
\caption{Decoupling temperature for process of fig.\ \ref{scatt} as a
function of gauge kinetic mixing parameter, for several values of the 
large
mass splitting, and for DM mass $M_\chi=5$ GeV
and gauge boson mass $\mu=100$ MeV.} \label{kindec}
\end{figure}

The principal interaction for maintaining kinetic equilibrium with
the SM is the electron scattering diagram shown in fig.\
\ref{scatt}.  The rate for this process is computed in appendix
\ref{kineticSM}.  The decoupling temperature as a function of
$\epsilon$ (the kinetic mixing parameter for whichever gauge boson is
exchanged) is shown in fig.\ \ref{kindec} for the case $M_\chi=5$ GeV
and $\mu=100$ MeV. This can easily be generalized to other DM and
gauge boson masses by noticing that the rate scales like $\alpha_g
\epsilon^2/\mu^4$ and $\alpha_g$ is proportional to $M_\chi$.  Hence
the scaling of $\epsilon$ in fig.\ \ref{kindec}.   For lower values of
$\epsilon$ than shown in the figure, the relation extrapolates to
a power law,
\beq
	{T\over 10{\rm\ MeV}}\cong \left(\epsilon\over 1.2\times 10^{-6}\right)^{-2/3}
\eeq

In reality there are two transitions with two different mass
splittings that can maintain kinetic equilibrium, since we also
have the $\chi_2 e\leftrightarrow \chi_3 e$ reaction with the small
mass splitting $\delta M_{23}$.  We compute the
decoupling temperature for both reactions and take the smaller of the
two as the true $T_d$.  Roughly speaking, only the larger of the
two $\epsilon$'s is therefore relevant for kinetic equilibrium.  
Figure \ref{kindec} shows that there is a weak dependence upon
$\delta M$ with the large mass gap giving a bigger effect.  
There is also dependence upon the gauge boson masses.

\begin{figure*}[t]
\smallskip \centerline{\epsfxsize=0.45\textwidth\epsfbox{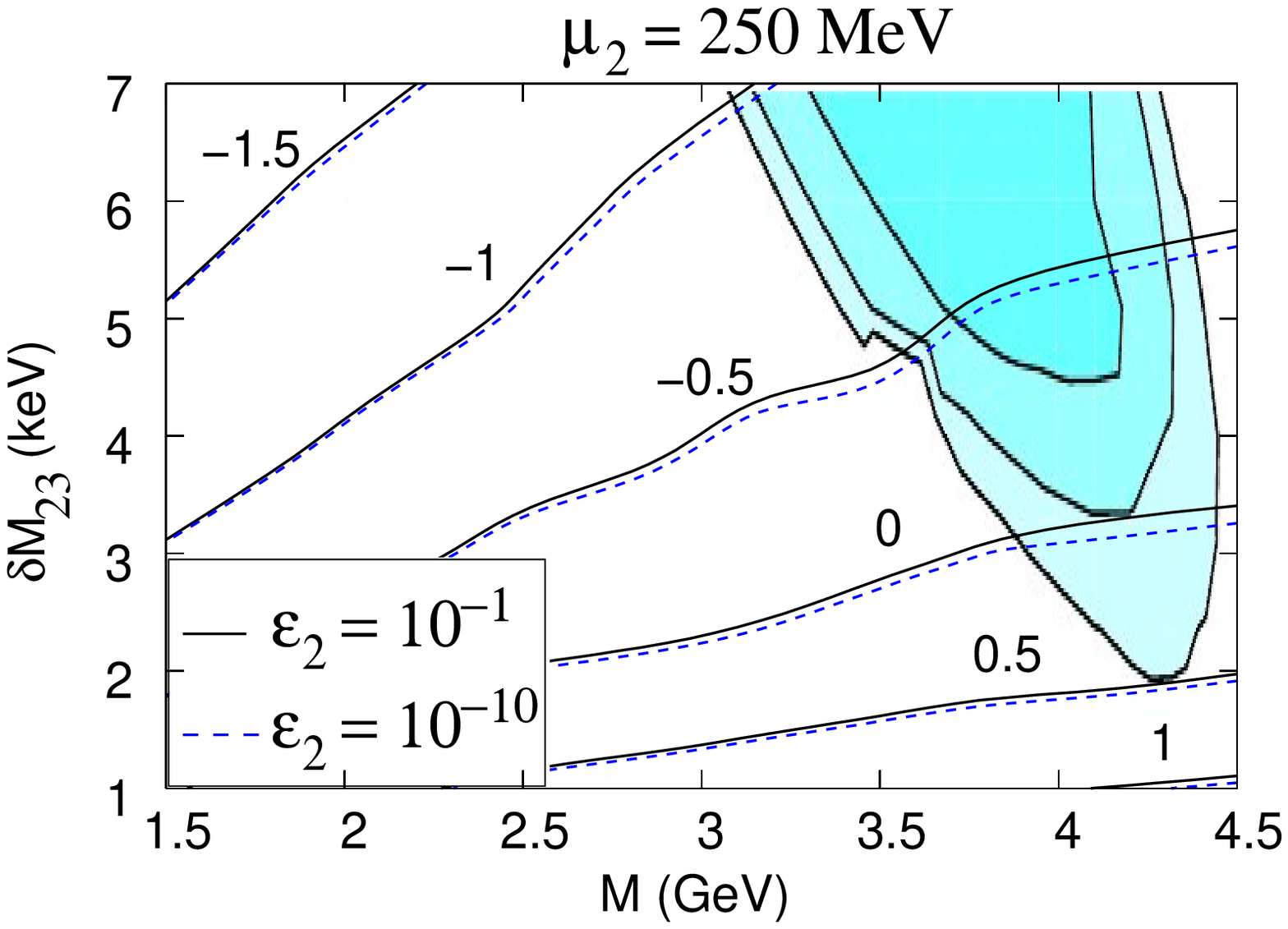}
\hfil\epsfxsize=0.45\textwidth\epsfbox{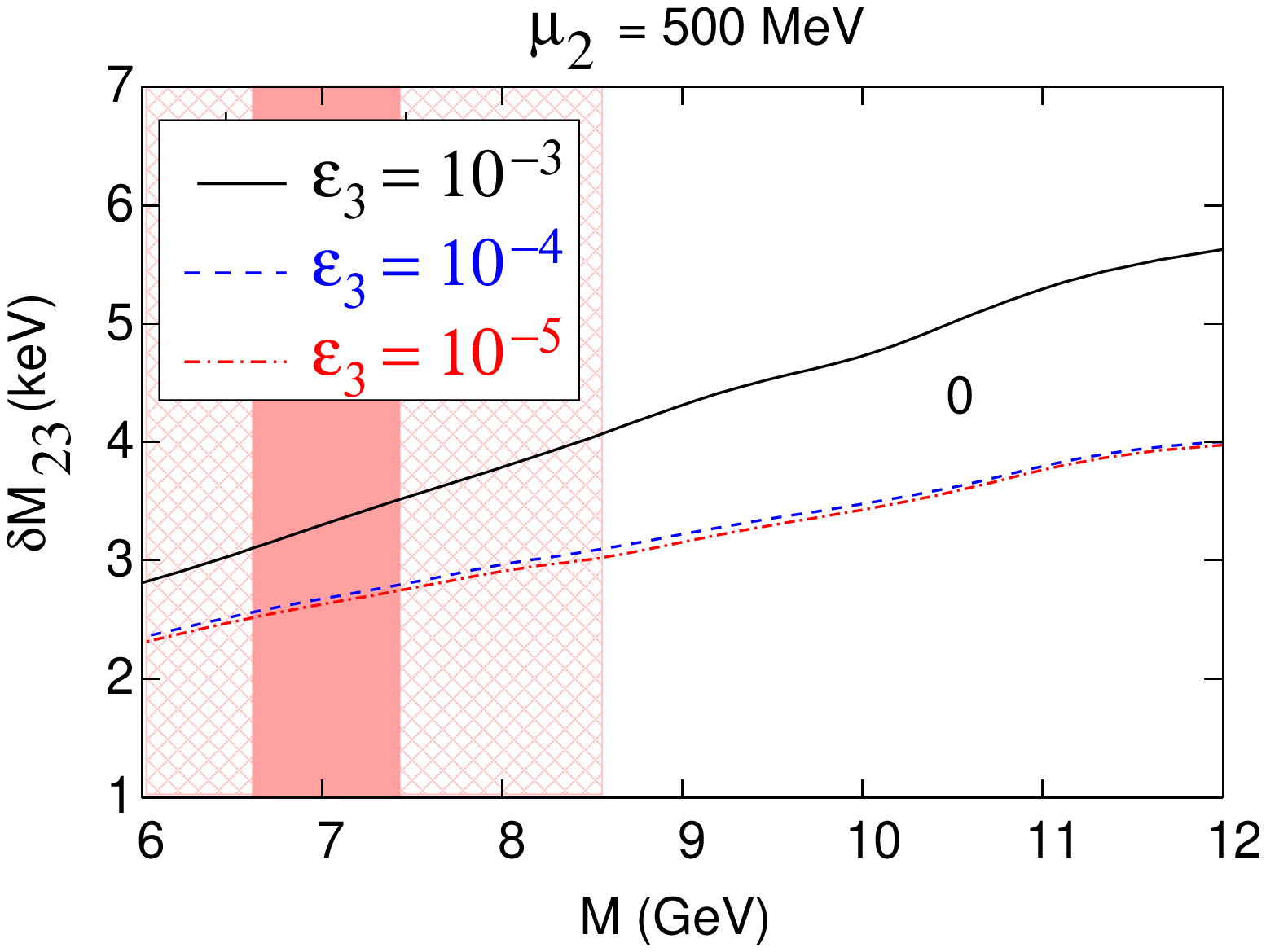}}
\caption{To illustrate the effect of $\epsilon_{2,3}$ on kinetic
decoupling and the relic density of the excited state, left: 
contours 
of $\log R_{e^+}/R_{\rm obs}$ in the exothermic model, varying
$\epsilon_2$ between $10^{-1}$ and $10^{-10}$.  $\mu_2$ is fixed at
250 MeV and other parameters are as in fig.\ \ref{exothermic}.
Right: similar plot for the endothermic model, with $\mu_2 = 500$
MeV.  For clarity only the contours with $\log R_{e^+}/R_{\rm obs}=0$
are shown.  Dependence on $\epsilon_3$ is saturated for
$\epsilon_3>10^{-3}$ or $\epsilon_3<10^{-5}$.}
\label{kindec2}
\end{figure*}

For the exothermic model, the coupling $\epsilon_1$ is large enough
so that $\epsilon_2$ is practically irrelevant for kinetic
equilibrium.  This is illustrated in figure \ref{kindec2} (left
panel), which shows that  contours of $\log R_{e^+}/R_{\rm obs}$ 
hardly change between $\epsilon_2= 10^{-1}$ and $10^{-10}$.
(The example shown is for $\mu_2 = 250$ MeV; for larger $\mu_2$ the
dependence is even weaker.)    
For the endothermic model, $\epsilon_1$
is smaller and so $\epsilon_3$ can have a bigger impact.  The 
right panel of fig.\   \ref{kindec2} shows that $\delta M_{23}$
must decrease by about 1 keV in the INTEGRAL/DAMA-allowed region 
to compensate
the effect of making $\epsilon_3$ arbitrarily small.

\section{Positron production rate and angular profile}
\label{rate-angle}

\subsection{Rate from inelastic collisions}
\label{rate}

The most recent determination of the observed positron annihilation
rate in the bulge is $1.1\times 10^{43}$/s \cite{bouchet2}.
This value depends upon the assumed distance between the sun and
the galactic center; consistently with  \cite{bouchet2} we take 
$r_\odot = 8.5$ kpc \cite{klb}. For the predicted rate, we have
\beq
	R_{e^+} = \frac12 \left(Y_s\over Y_{\rm tot}\right)^2\int d^{\,3} x \, \langle \sigma v\rangle\, {\rho^2\over
	M_\chi^2}
\label{posrate}
\eeq
where $\sigma$ is the cross section for $\chi_s\chi_s\to \chi_u\chi_u$
(recall that $\chi_{s,u}$ are the stable and unstable excited states).
The  $1/2$ is to avoid double-counting, and the abundance
factor $Y_s/Y_{\rm tot}$ is given by (\ref{Ys}).
We integrate over a region of radius 1.5 kpc, corresponding to 
an angular diameter of approximately 10$^\circ$.  The 
observed profile, fig.\ \ref{angular} suggests that the signal 
falls below the sensitivity of INTEGRAL near this angle.

The phase space average of $\sigma v$ is given by
\beq
	\langle\sigma v\rangle = \int d^{\,3}v_1 d^{\,3}v_2\, f(v_1)\, f(v_2)
	\sigma |\vec v_1 - \vec v_2|
\eeq
We take a Maxwellian velocity distribution 
\beq
	f(v) = N e^{-v^2/v_0^2}\,\theta(v-v_{\rm esc})
\eeq
cut off at the escape velocity
\beq
	v_{\rm esc}^2(r) = 2 v_0^2(r)\left[2.39 + 
\ln({\rm 10\ kpc}/r)\right]
\label{vesc}
\eeq
and having velocity dispersion
\beq
	v_0(r)^3 \propto r^\chi\,\rho(r)
\label{bert}
\eeq
with $\chi = 1.64$, and the normalization such that $v_0(r_\odot)$ is
220-230 km/s.  This form of $v_0$ is suggested by $N$-body simulations
that include the effects of baryonic contraction \cite{tissera}.  Our
choice of $v_{\rm esc}$ follows ref.\ \cite{CC}; see appendix C of that
paper.
  
In our previous work, the major challenge was to compute
$\sigma$ since we were concerned with DM at the TeV scale, implying
gauge  couplings $\alpha_g$ larger than the average DM velocity $v$. 
In this case a nonperturbative calculation of $\sigma$ was necessary,
since multiple gauge boson exchanges occur when $v < \alpha_g$,
similarly to the Sommerfeld enhancement in DM annihilation
\cite{sommerfeld}.  However
in the present situation $\alpha_g \ll v$ and a perturbative
treatment suffices.  

We define some kinematic variables to facilitate the presentation of the
cross section:
\beq
	v_t^2 = 2{\delta M_{23}\over M_\chi}, \qquad \Delta = {v^2\over
	v_t^2}
\eeq
where $\delta M_{23}$ is the small splitting between the two excited states,
$v$ is the DM velocity in the center of mass frame, and $v_t$ is the
threshold velocity for $\chi_2\chi_2\to \chi_3\chi_3$ excitations.
The cross section for excitations can be expressed as
\beq
	\sigma_{\uparrow} v_{\rm rel} = 4\pi\alpha_g^2
	\sqrt{\Delta -1}
	{ v_t  M_\chi^2\over {\cal D}^2} \left({2\over 1-\eta^2}-{1\over 2\eta}
	\ln{1+\eta\over 1-\eta}\right)
\eeq
where
\beq
	{\cal D} = M^2_\chi v_t^2 (2 \Delta -1) + \mu_1^2;
	\qquad\eta = 2 {M^2_\chi v_t^2\sqrt{\Delta(\Delta-1)}\over 
	{\cal D}}
\eeq
Notice that $\sigma_{\uparrow} v_{\rm rel}$ vanishes at threshold, $\Delta = 1$.  The related cross
section for  $\chi_3\chi_3\to \chi_2\chi_2$ de-excitations is
\beq\label{sigma_down_full}
	\sigma_{\downarrow} v_{\rm rel} = 4\pi\alpha_g^2
	\sqrt{\Delta +1}
	{ v_t  M_\chi^2\over \bar{\cal D}^2} 
	\left({2\over 1-\bar\eta^2}-
	{1\over 2\bar\eta}
	\ln{1+\bar\eta\over 1-\bar\eta}\right)
\eeq
where
\beq
	\bar{\cal D} = M^2_\chi v_t^2 (2 \Delta +1) + \mu_1^2;
	\qquad\bar\eta = 2 {M^2_\chi v_t^2\sqrt{\Delta(\Delta+1)}
	\over \bar{\cal D}}
\eeq
As expected, $\sigma_{\downarrow} v_{\rm rel} $ can be obtained
from $\sigma_{\uparrow} v_{\rm rel}$ by changing $\delta M_{23}\to
-\delta M_{23}$, which implies $v_t^2\to -v_t^2$ and $\Delta\to
-\Delta$ (notice that  $v_t\sqrt{\dots} = \sqrt{v_t^2\dots}$).
In the limit $\Delta\to 0$, and substituting $\delta M_{23}\to \delta
M_{1s}$ and $\mu_1\to\mu_i$, we recover the zero-velocity cross section 
for downscattering through the large mass gap, (\ref{sigma_down}). 

\subsection{Rate from decaying DM}
\label{ddmrate}

We consider the scenario where the ``unstable'' state
$\chi_u$ is so long lived that it is already present in the galaxy
due to its relic density, and decays with a lifetime $\tau_u$ greater
than the age of the universe.  Assuming that the 511 keV gamma rays observed
by INTEGRAL come from a central region of radius $r_c$, the rate of
positrons is 
\beqa
	R_{e^+} &=& {4\pi \over M_\chi\, \tau_u } \int_0^{r_c} dr\,
	r^2 \rho_u(r) \nonumber\\
	&\equiv& 4\pi\,\zeta\, {\rho_\odot\, {\rm kpc}^3 \over M_\chi\, \tau_u}
\left(Y_u\over Y_{\rm tot}\right)
\label{taudecay}
\eeqa
where  $\rho_u$ is the density of  
$\chi_u$ and $Y_u/Y_{\rm tot}$ is the abundance of $\chi_u$ relative to the total
DM population.  We assume the Einasto profile to obtain the
dimensionless factor
\beqa
\zeta &=&   
\left({r_s\over {\rm kpc}}\right)^3 e^{(2/\alpha)(r_\odot/r_s)^\alpha}
	 \nonumber\\
&\times& {1 \over \alpha} \left({\alpha\over
2}\right)^{3\over\alpha}\left[\Gamma\left({3\over\alpha}\right)-
	\Gamma\left({3\over\alpha},{2\over\alpha}\left({r_c\over
r_s}\right)^\alpha\right)\right]
\label{zetaeq}
\eeqa

\begin{figure}[t]
\smallskip \centerline{\epsfxsize=0.4\textwidth\epsfbox{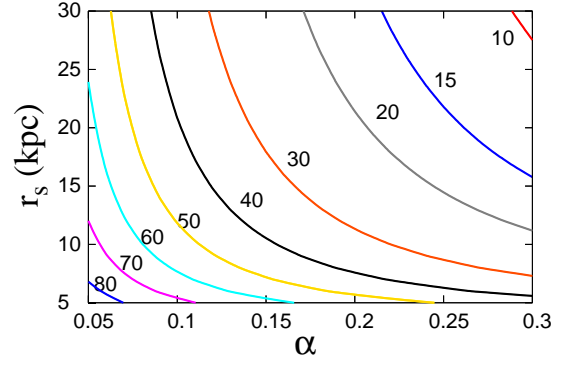}}
\caption{Contours of $\zeta$, defined in (\ref{zetaeq}), in the plane
of the Einasto halo parameters, with $r_c/r_\odot = 0.176$.  
$\zeta$ is related
to the volume integral of the DM density in the region of the INTEGRAL
511 keV signal, eq.\ (\ref{repeq}). } \label{zeta}
\end{figure}

Matching $R_{e^+}$ to the observed rate $3.4\times 10^{42}$/s, we find
that the lifetime of $\chi_u$ relative to the age of the universe 
(here we define $\tau_U\equiv 10^{10}{\rm y}$ rather than the actual
age of the universe)
\beq
	{\tau_u\over \tau_U} = 2.1\times 10^5\, 
	\left({\zeta\over 30}\right) \left(3 Y_u\over Y_{\rm tot}\right) 
	\left({5{\rm\ GeV}\over M_\chi}\right) 
	\left({\rho_\odot\over\bar\rho_{\odot} }\right)
\label{repeq}
\eeq
where $\bar\rho_{\odot}=0.3\,{\rm GeV/cm}^3$.
The factor $\zeta$ is plotted over a wide range of
Einasto parameters in fig.\ \ref{zeta},  showing that it is between
25 and 75 for reasonable profiles.  With $Y_u/Y_{\rm tot}\sim 1/3$
this gives $\tau_u \sim  10^{5}$ times 
$\tau_U$.

To see how small $\epsilon_{2,3}$ this corresponds to, we can 
rescale the bound   (\ref{epsbound}) to be derived below from demanding that similar
decays of the  ``stable'' excited state must take
longer than $\tau_U$.  It implies that 
\beqa
	\epsilon_{2,3}^2 &\cong& (7\times 10^{-12} )^2
 \left({10^{-4}\over \alpha_g}
	\right)\left({\mu_{2,3}\over 1\ {\rm GeV}}\right)^4
	\left({0.1\ {\rm MeV}\over \delta
M_{1s-}}\right)^{3}\nonumber\\&&\times
\left(50\over\zeta\right) \left(Y_{\rm tot}\over 3Y_u\right)
	\left({ M_\chi\over 5{\rm\ GeV}}\right)
	\left({\bar\rho_\odot\over\rho_{\odot} }\right) 
\label{eps23eq}
\eeqa
where $\delta
M_{1s-} = \delta M_{1s}-2 m_e$ is the energy available for the decay.
It is theoretically easy to achieve the desired rate of positron
creation just by
adjusting $\epsilon_{2,3}$ to this small value, since there is no
other constraint on $\epsilon_{2,3}$.

\subsection{Angular distribution} 

In this section we elaborate on the angular profile of the 511 keV signal
in the case of scatterings only, since only there is it definitely
necessary to consider the effects of positron propagation.
The intensity of the signal as a function of angle is found by computing
the line-of-sight integral (\ref{los})
where the line is oriented along the direction $\hat x$ specified by 
angles $\theta,\phi$ relative to the galactic center.  This expression
assumes that the positrons decay at the same position where they were
created.  To model the effects of propagation before decay, we smear
the angular distribution by averaging $\hat x$ weighted by some 
function $f(\cos \theta)$,
\beq
	\bar I_{e^+}(\hat x') = 
	\int d\Omega f(\hat x\cdot \hat x') I_{e^+}(\hat x)
\eeq
The integral over solid angle can be combined with the integral over the
line of sight and rewritten in terms of a volume integral,
$dx \, d\Omega = d^3x/x^2$, with the origin of coordinates at the
earth.  Now,
given that $\rho^2$ is strongly peaked near
the galactic center,  we can write $\rho^2 \sim \delta^{(3)}(\vec x-
\vec x_0)$, where $x_0$ is the position of the galactic center.  Then
we find that
\beq
	\bar I_{e^+}(\theta) \sim f(\cos\theta)
\eeq
The intensity has the same shape as the smoothing function.  As argued in 
section \ref{res-angular}, this is a good approximation for the DM halo
profiles that we are considering for the inelastic scattering
mechanism.

It is interesting to notice that even though $\rho^2$ looks like a delta
function with respect to the measure $d^3x/x^2$, not so for the usual
volume measure $d^3x$.  Indeed, the function $r^2 \rho^2(r)$ has a maximum
near $r= \sqrt{2} r_s$ even in the limit $\alpha\to 0$.  Therefore the
total rate of positron production in the galaxy gets significant
contributions away from the galactic center, although these are not 
counted in the observations of the bulge component upon which we focus in this
paper, since only near the center is the intensity high enough to be
detected.

\subsection{Regions of positron annihilation}

In the above discussion we have assumed that positrons  are able to
migrate to the regions where positronium forms and where they can
subsequently annihilate.  It is known from fitting the observed
$\gamma$ ray spectrum that $\sim 97$\% of the positrons indeed form
positronium before annihilating \cite{Churazov,spectral}.
This is because orthopositronium decays to three photons, and
comparison of the 511 keV line flux with the continuum level is
consistent with nearly all annihilations coming from positronium
rather than positrons encountering free electrons.  The spectral shape
also shows that most annihilations take place in warm ($\sim 8000$ K)
\cite{guessoum} regions, which may be mostly ionized \cite{Churazov}
or else a combination of neutral and ionized regions \cite{spectral}.

\begin{figure}[t]
\smallskip \centerline{
\epsfxsize=0.5\textwidth\epsfbox{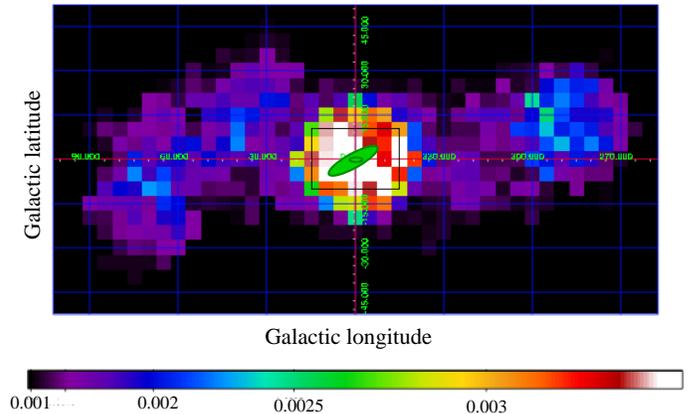}\hfil}
\caption{Superposition of molecular gas regions of \cite{Ferriere} (green ellipses in
center) on the intensity map of the INTEGRAL 511 keV observations
from ref.\ \cite{bouchet2}.  Innermost ellipse is the CMZ (central
molecular zone); outer tilted ellipse is the holed galactic bulge
disk.}
\label{int-int}
\end{figure}

Efforts have been made to independently map out the positions of 
the warm regions in the galactic bulge (GB); doing so could provide a
consistency check on the above determinations, since then the
morphology of the INTEGRAL detection of the galactic bulge 511 keV
gamma rays should match the position of the warm regions.  Ref.\ 
\cite{Ferriere} has modeled the spatial distribution of 
molecular gas in the GB based on CO emission data \cite{sawada}
for the central molecular region (CMZ) in the inner 150 pc, and
borrowing an older model \cite{burton} based on H{\sc i} observations
for the ``holed GB disk'' region extending to radii of $\sim 1$ kpc.
It has been suggested that the warm neutral or ionized regions
relevant for positronium ionization coincide with these molecular
gas clouds \cite{higdon}.\footnote{Ref.\ \cite{higdon} assumes that
positrons from the radioactive decays of supernova ejecta 
can be transported from the galactic disk into the GB to account for 
the observed 511 keV signal; however the validity of their model
of electron transport has been questioned \cite{Jean,Skinner}.}\ \ 
In figure \ref{int-int} we have tranposed
an outline of the CMZ and holed disk regions (fig.\ 4 of ref.\
\cite{Ferriere}) on the most recent INTEGRAL 511 keV intensity map
\cite{bouchet2}.

From fig.\ \ref{int-int}, it is clear that the INTEGRAL/SPI
instrument does not have sufficient spatial resolution to 
test whether positron annihilation really comes from the molecular
gas regions.  Furthermore, the assumption that these regions coincide
with the warm ionized or neutral $+$ ionized regions of positronium
annihilation is questionable.   The direct measurements of the ionized
component are based upon pulsar observations  \cite{CL}, which
suggest the existence of warm H$^+$ regions with similar morphology
to the molecular gas.  But this is not considered to be  a very
reliable measurement of the ionized gas density in the GB due to the
scarcity of pulsars in this region \cite{vicky}.  
Thus we do not know with a high level of confidence where the
warm regions of positronium annihilation are really located.  It is
possible that they extend beyond the molecular gas regions identified
by \cite{Ferriere}.

Because of the lack of very reliable information as to the spatial
distribution of the warm ionized ($+$ neutral) regions, an
uncertainty that is acknowledged in careful studies such as
\cite{Ferriere}, it is  possible that decays of DM lead to
positronium production in the vicinity of the initially produced
positrons, so that the INTEGRAL signal could be a reflection of the
underlying DM distribution.  On the other hand if the positrons
result from DM scattering, we have shown that they are initially
produced within $1^\circ\sim 150$ pc of the galactic center (dotted
red curve of fig.\ \ref{angular}), and then transport of the positrons to
larger radii is probably necessary to be consistent with the observed
extent of the 511 keV signal.  Positron transport in the galactic
center has been extensively studied, and shown to depend sensitively
on the largely unconstrained nature of the magnetic fields in this
region.

Ref.\ \cite{Jean} recently showed that, under reasonable
assumptions, positrons can travel well outside of the GC before
annihilating.   If this is the case, the initially highly localized
source from DM scattering will be widened to fill the interstellar
medium.  More information will be needed to attach a firm
interpretation to the angular distributions of the DM decays or
annihilations.  Observations of the 511 keV gamma rays using a future
instrument with better spatial resolution would clearly be
desirable for helping to settle these questions.  In particular, if
a new measurement revealed stronger localization of the GB component
of the 511 keV emission toward the galactic center, it would favor
the DM explanation over astrophysical sources.

\section{Direct detection rates}
\label{ddrates}

In our computation of the 511 keV rate, we fixed the value of
gauge kinetic mixing parameter $\epsilon_1$ so as to match the
direct detection rates determined respectively by references
\cite{harnik} and \cite{hooper} for the exothermic and endothermic DM
models.  Although $\epsilon_1$ does not directly affect the rate of
$\chi_2\chi_2\leftrightarrow\chi_3\chi_3$ transitions, it does so
indirectly, because of its influence (through kinetic decoupling)
on the relic density of the stable excited state.  Here we give
details on the determination of $\epsilon_1$ in these two cases.  We note
briefly that the nuclear form factor is trivial for collisions studied
here, so we will ignore it.

\subsection{Exothermic dark matter}

Ref.\ \cite{harnik} determined the elastic limit of the
DM-nucleon cross section needed to
get the right rate of DAMA transitions:
\beq
	\sigma_{n,\rm el} = {\mu_n^2\over {4\pi\Lambda^4}}\ ,
\label{sigma_harnik}
\eeq 
where $\mu_n = m_n M_\chi/(m_n+M_\chi)$ is the reduced mass and 
$\Lambda = 340$ GeV.\footnote{Note that taking this elastic limit negates
the need to average over DM speeds as in (\ref{sigma_endo}).}  
In our model, the coupling is to protons only,
and the analogous quantity  is given by
\beq
	\sigma_{p,\rm el} = 16 \pi \epsilon_1^2 \alpha\alpha_g
	{\mu_n^2\over \mu_1^4}\ .
\label{sigma_pel}
\eeq
To determine the value of $\epsilon_1$ needed to match the observed
rates, we must account for the coupling to protons only since
the rate is proportional to $A(Z f_p + (A-Z)f_n)^2$ for atomic
number and mass $Z,A$ and relative strengths of couplings to protons
and neutrons $f_p,f_n$. In ref.\ \cite{harnik}, the couplings were
assumed to be $f_p=f_n=1$, but we have $f_p=1$, $f_n=0$. 
Moreover, we have a different 
local density of the excited state than that assumed by \cite{harnik}
because of the abundance factor $Y_s/3Y_1$ (which also appears in the 
positron rate (\ref{posrate})), and because we allow the local DM
density to vary with respect to the fiducial value $\rho_0 = 0.3$
GeV/cm$^3$.  The result is
\beq
	\epsilon_1 = {\mu_1^2\over 8\pi\Lambda^2}{A\over Z}\left[
	{1\over \alpha\alpha_g}\,{Y_{\rm tot}\over Y_s}\,
	{\rho_0\over\rho_\odot}\right]^{1/2}
\label{eps1eq}	
\eeq
If $\epsilon_{2,3}\ll\epsilon_1$ so that $\epsilon_1$ determines the
kinetic decoupling temperature of the DM, then $Y_s$ depends
implicitly on $\epsilon_1$ and (\ref{eps1eq}) must be solved 
numerically.  The factor with $A/Z$ depends upon which nucleus we
are talking about, and is given by 2.4 and 2.28 respectively
for I and Na.  As \cite{harnik} notes, scattering from Na nuclei is 
preferentially detected in our region of parameter space, so we choose
the latter number.

\begin{figure*}[t]
\smallskip \centerline{
\epsfxsize=0.45\textwidth\epsfbox{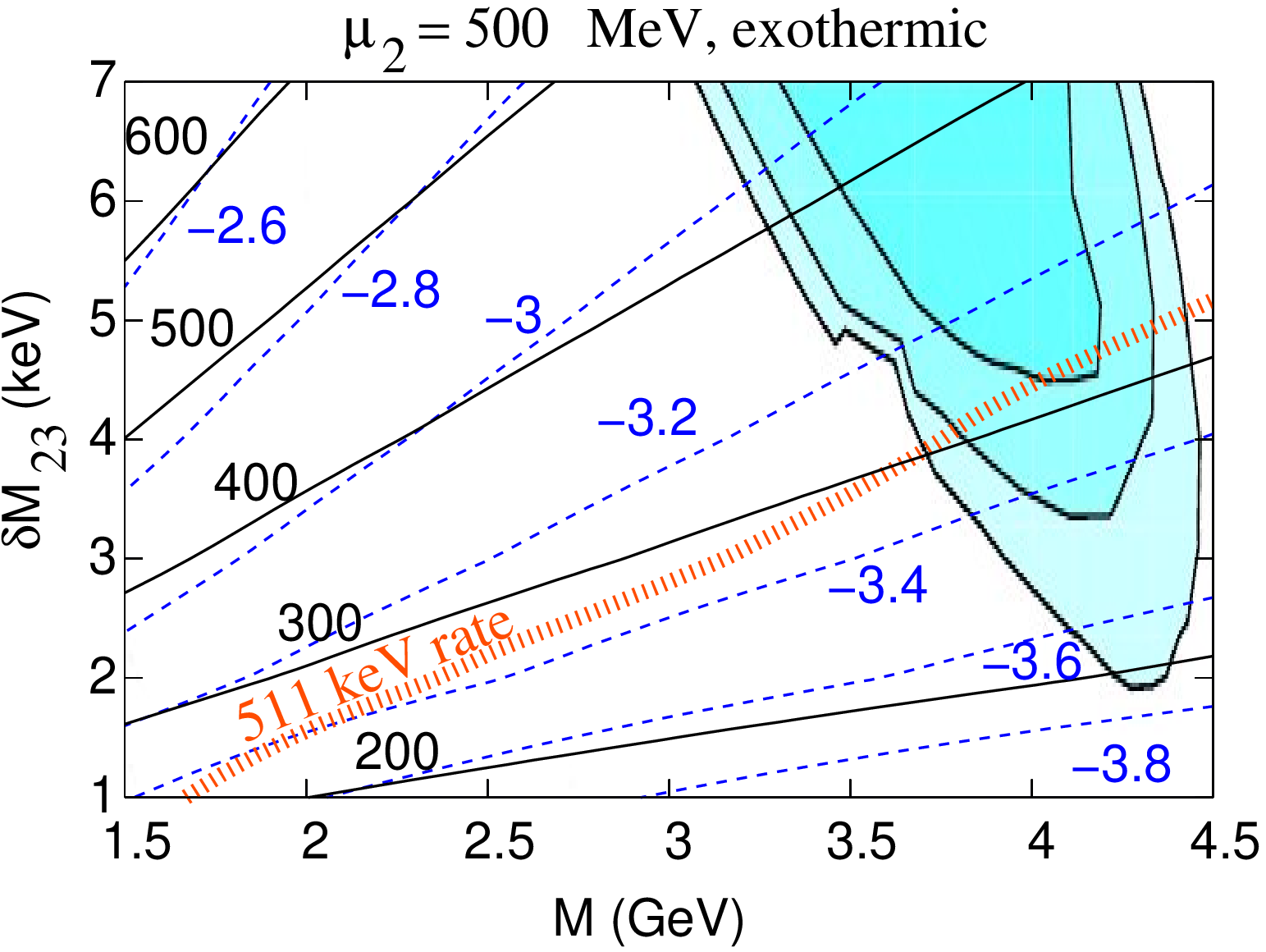}\hfil
\epsfxsize=0.45\textwidth\epsfbox{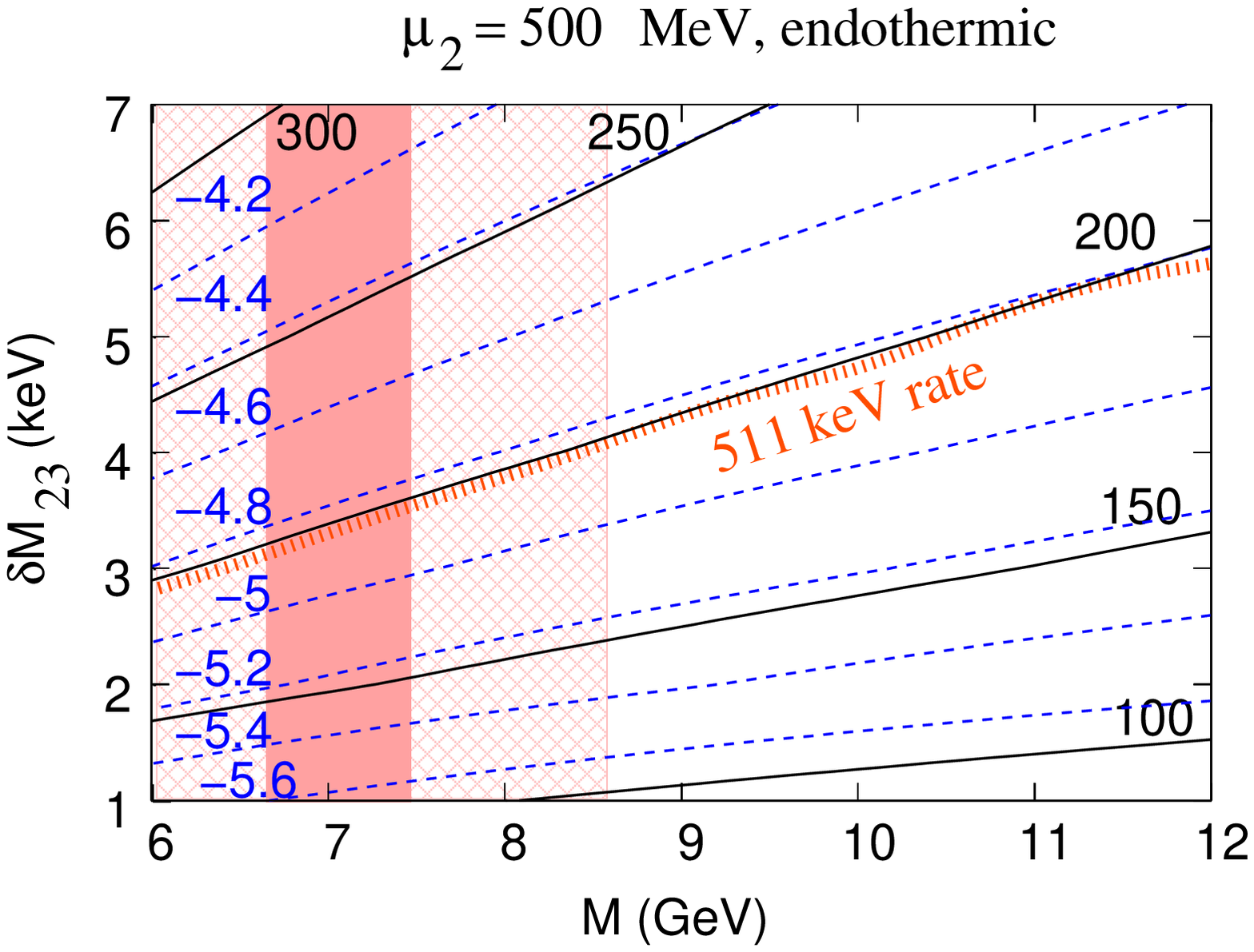}}
\caption{Contours of $\log\epsilon_1$ (dashed) and $\mu_1$ (solid,
in MeV) for $\mu_2=500$ MeV, in the exothermic (left) and endothermic
(right) models.  Thick curve labeled ``511 keV rate'' is the contour
where the predicted positron rate matches the INTEGRAL observation.
Shaded areas are the DAMA-allowed regions.}
\label{eps-mu}
\end{figure*}

In figure \ref{eps-mu} (left panel) we plot contours of $\epsilon_1$
corresponding to the $\mu_2=500$ MeV example shown in fig.\ 
\ref{exothermic}, to give a sense for how large $\epsilon_1$ must be.
Near $M_\chi = 4$ GeV, $\delta M_{23}= 4.5$ keV, where 
the INTEGRAL and DAMA rates best fit simultaneously, $\epsilon_1\sim
10^{-3.27}$, significantly larger than the generic estimate $10^{-5}$
given in ref.\ \cite{harnik}.   This is due to the $A/Z$
correction,  the fact that we need $\mu_1$ to be heavier
than the nominal 100 MeV value assumed in \cite{harnik}, and that
$Y_s/ Y_{\rm tot}$ can be significantly less than 1 in our
model.

\subsection{Endothermic dark matter}

Ref.\ \cite{hooper} finds that DM with a mass of approximately 7 GeV
and cross section on nucleons  
\beq
	\sigma_n = 2\times 10^{-4}{\rm\ pb}
\label{sigma_hooper}
\eeq
can
simultaneously fit the DAMA/LIBRA annual modulation and the CoGeNT
low-recoil events. Their allowed regions of $\sigma_n$ versus 
$M_{\chi}$  are reproduced in figure \ref{hooper_fig}. 
The logic for matching our cross section to theirs is similar
to the exothermic case, except for the
fact that endothermic scatterings are kinematically blocked 
if the DM velocity is below the threshold value
\beq
	v_t = \sqrt{2\delta M_{23}/\mu_N}
\eeq
where $\mu_N = m_N M_\chi/(m_N+M_\chi)$ is the 
nucleus-DM reduced mass.  We take this into
account by doing the phase space average of 
$\sigma v$.  
The phase space factor in $\sigma v$ that is sensitive to the
threshold is $\sqrt{v^2 - v_t^2}$.  For elastic scattering, this
factor would be $v$.  Therefore we match the quantity 
(\ref{sigma_pel}) that also appears in 
our slightly inelastic cross section to (\ref{sigma_hooper}) using
\beq\label{sigma_endo}
	\sigma_{p,{\rm el}} = {\langle v\rangle \over 
	\langle \sqrt{v^2 - v_t^2}\,\rangle} 
	\left({A\over Z}\right)^2
{\rho_0\over\rho_\odot}\, {3Y_1\over Y_s}\,\sigma_n\ ,
\eeq
where the averages over velocity are performed with the  
Maxwellian distribution function $f= N
e^{-v^2/v_0^2}$ cut off at the escape velocity $v_{\rm esc}$.  Since
we are comparing with  ref.\ \cite{hooper}, we use their values
$v_0=230$ km/s and $v_{\rm esc} = 600$ km/s for this part.  Once again,
scatterings from Na are preferentially detected, so $A/Z=2.28$.
Similarly, we take the threshold velocity for sodium in the above.


Figure \ref{eps-mu} (right panel) shows contours of $\epsilon_1$ 
for the case of $\mu_2 = 500$ MeV.  In the overlap region for
INTEGRAL and DAMA, $\epsilon_1\cong 10^{-5}$.  This is smaller
than required in the exothermic model because the corresponding
value of $\mu_1$ is smaller, and also the cross section 
(\ref{sigma_hooper}) is approximately 0.15 that in (\ref{sigma_harnik}).
Kinetic equilibrium of the DM with the SM in the early universe
is not as efficiently maintained by $\chi_2\leftrightarrow\chi_3$
transitions in this case.  This is why the relic density of $\chi_2$
is sensitive to the value of the other nonvanishing kinetic mixing
parameter for the endothermic model, whereas it is practically 
insensitive in the exothermic case.

\section{Astrophysical constraints}
\label{constraints}

In this section we address the astrophysical and cosmological  constraints on 
our proposal that are complementary to the 511 keV and direct
detection signals, as well as to laboratory constraints from 
electron beams.

\subsection{Lifetime of metastable state}\label{life_metastable}

We need to insure that the ``stable'' excited state $\chi_s$  is
either truly stable or else sufficiently long-lived.  The most
dangerous process is $\chi_s\to \chi_1 e^+ e^-$.  
At the phenomenological level,
we suppress this by setting the kinetic mixing of the gauge
boson that mediates this process to a sufficiently small value.
The rate for this decay can be computed analytically with the
approximation that $\delta M_{1s-}=\delta M_{1s} - 2 m_e$ is sufficiently small
for the final state particles to be nonrelativistic.  Then
\beq
	\Gamma_{\chi_s} \cong 2\alpha_g\alpha\epsilon^2
{m_e^2\, \delta M_{1s-}^3\, \mu^{-4}}
\label{chisdecay}
\eeq
\\ 
Demanding that $\tau_{s}$ exceed $10^{10}$ y requires that
\beq
	\epsilon < 2\times 10^{-9} \left({10^{-4}\over \alpha_g}
	\right)^{1/2}\left({\mu\over 1\ {\rm GeV}}\right)^2
	\left({0.1\ {\rm MeV}\over \delta M_{1s-}}\right)^{3/2}
\label{epsbound}
\eeq

\begin{figure}[h]
\includegraphics[scale=0.25]{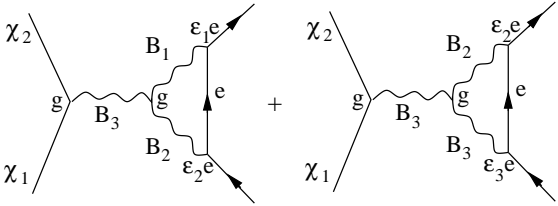}
\caption{\label{chidecay} Decay of metastable $\chi_2$ state
due to nonzero $\epsilon_{1,2}$.  
}\end{figure}

In the endothermic model, we can set $\epsilon_3=0$ at tree level by 
removing the $\Delta_3$ Higgs boson, but
there seems to be no symmetry to ensure that $\epsilon_3$ is not
generated by loops 
if the other two mixing parameters are nonzero.  Nonetheless, 
we are not able to
find an example of a loop diagram that generates nonzero $\epsilon_3$;
any that superficially seem promising vanish because of Furry's
theorem.  Instead, we find a one-loop process where $B_3$ acquires
a magnetic moment coupling to the electron, $\mu_B B^{\mu\nu}\bar e
\sigma_{\mu\nu} e$.  
The decay of $\chi_2$
proceeds by $B_3$ exchange in the one-loop diagrams of fig.\ 
\ref{chidecay}.  These diagrams would cancel exactly if
$\mu_2 = \mu_1$, so the magnetic moment can be estimated as
\beq
	\mu_B\sim g \alpha \epsilon_1 \epsilon_2\,{
	(\mu_2-\mu_1)
	\over 4\pi\,\bar\mu^2}\ln {\Lambda\over \bar\mu}
\eeq
where $\bar\mu = \frac12 (\mu_1+\mu_2)$ and $\Lambda$ denotes the 
hidden SU(2) symmetry breaking scale,
above which the kinetic mixing of $B_{1,2}$ is replaced by an
interaction with the triplet Higgs fields.  The squared matrix element
of fig.\ \ref{chidecay} can be estimated as $|{\cal M}|^2\sim g^2\mu_B^2 M_\chi^2 m_e^2
\delta M_{12}^2 \mu_3^{-4}$, and the decay rate in the limit of small
$\delta M_{12-} \equiv \delta M_{12} - 2 m_e$ is 
\beq
	\Gamma_{\chi_2\to \chi_1 e^+e^-}\ \sim \
	{\alpha_g \over 32\pi} \mu_B^2\, m_e^3\, \delta M_{12}^2\,
	\delta M_{12-}^2\, \mu_3^{-4}\label{loopdecay1}
\eeq
For $\alpha_g\sim
10^{-4}$ and $\epsilon_1\sim \epsilon_2\sim 10^{-3}$,
$\delta\mu\sim\mu_3 \sim
\bar\mu\sim 1$ GeV, $\delta M_{12-}\sim 0.1$ MeV,
$\Lambda \sim 10$ GeV, we find a lifetime of $10^{26}$ s, much larger
than the age of the universe.  
 Therefore it seems technically natural
to neglect the dangerous kinetic mixing term and assume the ``stable''
state is sufficiently long-lived.  As it turns out, a careful
calculation is even more suppressed; see appendix \ref{loopdecay}.

The $\chi_3\to\chi_2 X$ decay in the exothermic model is not
problematic, since the mass splitting is much smaller and the only
available decay channels are with $X=\gamma$, $X=\nu\bar\nu$, and
$X=3\gamma$.  
These have been studied previously \cite{nonabelian,Batell-multicomponent,
Finkbeiner-metastable}.  The single-photon decay has a lifetime longer
than the age of the universe for the value of $\epsilon_1$ required for
direct detection; since it could be observed, we discuss it in more detail
in section \ref{spdms} below.  

The partial width
for the $\nu\bar\nu$ final state is easy to estimate in  analogy with
(\ref{chisdecay}).  For this channel, 
there is an additional suppession in the kinetic mixing.  First, 
the mixing of $B_1$ with the $Z$ boson current has an extra factor of
$\mu_1^2/m_Z^2$ \cite{nonabelian}, and, second, the SM $Z$ boson mixes 
with the $B_1$ current with opposite sign such that the two mixings
nearly cancel at small energy-momentum transfer (see appendix 
\ref{kineticSM} for discussion of the same cancellation in 
$\chi\nu$ scattering).  
The ensuing bound on $\epsilon_1$ is much weaker than
that on $\epsilon_2$; 
practically speaking there is no constraint.

The decay $\chi_3\to\chi_2 +3\gamma$ is due to the operator $\sim 
(\epsilon \alpha^2/90 m_e^4)
B_1^{\mu\nu}F_{\mu\nu} F^2$ induced by an electron loop, similar
to the Euler-Heisenberg $F^4$ interaction in QED.  (Furry's theorem
forbids a term of the form $B_1 F^2$ and $B_1 F$ mixing is already
taken into account by diagonalizing the kinetic terms.)  The
rate is suppressed by $\delta M_{23}^{13}$
\cite{Finkbeiner-metastable,
Batell-multicomponent}, leading to lifetimes that
far exceed the age of the universe for the small $\sim 5$ keV
splittings relevant to our exothermic model.

\subsection{Single-photon decays of metastable state}
\label{spdms}

The exothermic version of our proposal faces the challenge that
the excited state can decay by emission of a single photon,
via $\chi_3\to\chi_2\gamma$.  The origin of this decay was pointed
out in \cite{nonabelian}: the nonabelian term in
the field strength $B_1^{\mu\nu}$ leads to interactions of the form
$\epsilon_{1} B_2^\mu B_3^\nu F_{\mu\nu}$ with the photon, from the
gauge kinetic mixing operators (\ref{kinmix}).  This can be put into 
a loop diagram which results in a transition magnetic moment
$\chi_2$-$\chi_3$,
\beq
	\mu_{23} \cong {\epsilon_1 g^2
\over
	128\pi^2 M_\chi}\left(\ln{M_\chi\over\mu}-1\right)\ ,
\label{mu23}
\eeq
where $\mu$ is of order 
$\mu_{2},\mu_{3}$.  Therefore, there is a
decay channel $\chi_3\to\chi_2\gamma$.  The rate is 
\beq
	\Gamma_\gamma = {\mu_{23}^2\over 8\pi}(\delta M_{23})^2
\label{gamma23}
\eeq
For $\epsilon_1 \cong 10^{-5}$, $M_\chi\cong$ 4.5 GeV and $\delta
M_{23}\cong 5$ keV, the lifetime is $4\times 10^{19}$ s which is much
longer than the age of the universe.  However, this is not
necessarily enough because such photons could be visible in
astronomical searches.  

Of the various instruments that could be sensitive to low-energy
$\gamma$ rays, INTEGRAL/SPI comes the closest. Ref.\ \cite{teegarden}
gives limits on the intensity of gamma ray lines that could come from
such decays in the galaxy; however INTEGRAL's sensitivity cuts out
below 20 keV, making our scenario just out of reach.  Interestingly
limits on the diffuse gamma ray background put a $\delta
M_{23}$-dependent lower limit on
the partial lifetime times the mass \cite{yuksel-kistler} of 
approximately
\beq
\tau_\gamma M_\chi > 1\times 10^{20} \left(\frac{3Y_3}{Y_{\rm tot}}\right)
\left(\delta M_{23}\over 10{\rm\ keV}\right)^{1.2} {\rm\ GeV\  s} 
\eeq
for $\delta M_{23} > 10$ keV. Again because of INTEGRAL's energy
sensitivity, data is not given for lower photon energies. 
Nevertheless, extrapolating the bound to
$\delta M_{23} =5$ keV gives $4.3\times 10^{19}$ GeV s, which is not even
five times less than
our nominal value $1.8\times 10^{20}$ GeV s, assuming $M_\chi =
4.5$ GeV.  Therefore an instrument sensitive to these lower energies
might detect this low-energy photon, which is in the x-ray
part of the spectrum. 

\begin{figure}[t]
\includegraphics[scale=0.4]{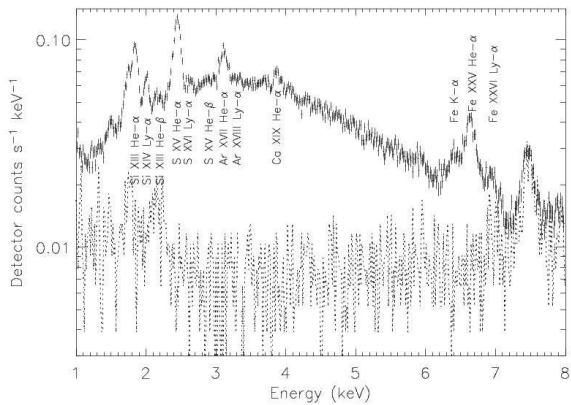}
\caption{\label{chandra-spect} Chandra spectrum from region near
galactic center, where $\sim 5$ keV x-ray from 
$\chi_3\to\chi_2\gamma$ decay might manifest itself.  
}\end{figure}

In fact, observations of the galactic center by the Chandra x-ray
telescope \cite{chandra} may rule out this  particular model.
Observations are presented for a region of size 35 arcmin$^2$ that is
7.5 arcmin away from the GC.  No evidence of an unidentified line is
observed in the $1-8$ keV band (fig.\ \ref{chandra-spect}), and the 
continuum seen there is modeled
by thermal sources with a flux of $6\times 10^{-12}$ erg cm$^{-2}$
s$^{-1}$.  We can compute the expected flux by integrating
over the line of sight and the solid angle ($d\Omega = d\phi\,
d\cos\psi$) of the observed region \cite{yuksel-kistler},
\beqa
F_{\rm th} &=& {Y_3/Y_{\rm tot}\over 4\pi M_\chi\tau_\gamma}
\int d\Omega \int dl\, \rho
(|\vec l-\vec r_\odot|)\nonumber\\
&\equiv & {\rho_\odot r_s\over 2 M_\chi\tau_\gamma} \frac{Y_3}{Y_{\rm tot}}I
\eeqa
where $I = e^{(2/\alpha)y^\alpha} \int d\cos\psi
\int d\hat l\, 	e^{-(2/\alpha)(y^2 + \hat l^2 - 2 \hat l y
\cos\psi)^{\alpha/2}}$, $y=r_\odot/r_s$ and $\hat l = l/r_s$.
We numerically integrate over an annular region of similar area and
displacement from the GC to the observed one,
using the Einasto parameters (\ref{einasto-params}) to find
$I\cong 10^{-4}$.  Using the value of $M_\chi\tau_\gamma$
determined above, this gives $F_{\rm th} \cong 0.001$ photons 
cm$^{-2}$ s$^{-1}$ (for the maximal ratio $Y_3/Y_{\rm tot}$.  
The corresponding energy flux for a 5 keV
mass difference is $10^{-11}$  erg cm$^{-2}$
s$^{-1}$, not quite 2 times greater than the observed continuum flux.  

In the above estimate we did not take account of absorption of the
decay signal, which would help to soften the discrepancy, especially
if the photon energy is somewhat lower.  It may also be possible to 
evade the problem by extending the gauge group to SU(2)$\times$U(1)
\cite{baum}
and replace the kinetic mixing of $B_1$ by that of the extra U(1)
gauge boson; this would remove the $\mu_{23}$ transition magnetic
moment.  Notice that this problem does not affect the endothermic
model because $\chi_3\to\chi_1e^+e^-$ proceeds much more quickly, as 
we discuss in section \ref{unstable} below.  

\subsection{Lifetime of unstable state}
\label{unstable}

In passing, we can also estimate the decay rate for the unstable
excited state into $e^+e^-$ using (\ref{chisdecay}).  It has the same
form, except for the substitutions of $\epsilon$ and $\mu$ by
the corresponding quantities for $B_3$, in the exothermic model; for
the endothermic model, (\ref{chisdecay}) applies as written, to the 
unstable excited state.  Laboratory experiments constrain the
appropriate $\epsilon_i$ to be $\lsim 10^{-3}$, so the lifetime could
be $10^{10}$ times shorter then the above estimate using
$\epsilon\sim 10^{-8}$, thus on order of
1 y.  This assumes the large mass splitting is only 1.1 MeV.  With a
2.1 MeV splitting one gains a factor of $10^5$ in the rate due to the
larger phase space, giving a lifetime of several hundred seconds.  It
cannot be significantly smaller in our model. 

\subsection{Single-photon decays of unstable state}

The decay mechanism discussed in section \ref{spdms} was originally
conceived for the decay of the unstable state in ref.\
\cite{nonabelian}.  This goes through the MeV-scale mass gap, so the 
photon in this case is a gamma ray.  
For definiteness let us consider the endothermic
model, so $\chi_3$ is the unstable state and the relevant decay
is $\chi_3\to \chi_1\gamma$ via the $\mu_{13}$ transition magnetic
moment, which is proportional to $\epsilon_2$, in analogy to
(\ref{mu23}).  The partial decay rate is the obvious generalization
of (\ref{gamma23}).  
 The branching ratio for the single photon decay relative to
that into $e^+e^-$ is \cite{nonabelian}
\beq
	{\rm BR}_\gamma = {\alpha_g^2 / \alpha\over 8192\pi^2}\,
	{\mu^4\, (\delta M_{13})^3\over M_\chi^2 
	(\delta M_{13-})^3(\delta M_{13+})^2} \,
	\ln^2{M_\chi\over e\mu}
\label{brg}
\eeq
where $\mu \sim \mu_1,\mu_3$, 
$\delta M_{13\pm}= \delta M_{13}\pm 2 m_e$ and 
$e = 2.71828\dots$.  The resulting photon might be observed
by INTEGRAL in the diffuse $\gamma$ ray background.
In ref.\ \cite{nonabelian}, a bound was derived, which however
overestimated the sensitivity of INTEGRAL to the signal.  We therefore
reconsider it here.

The analysis of ref.\ \cite{teegarden} is particularly relevant for
us, since they searched for line sources from the galactic center
region, having a spatial distribution
similar to that of the 511 keV line.  They limit the flux of such a
line, for energies between 1 and 2 MeV, to less than  $\sim 3\times 10^{-5}$
cm$^{-2}$ s$^{-1}$.  This is to be compared to the flux from
positrons, $\sim 3.6\times 10^{-3}$ cm$^{-2}$ s$^{-1}$.  Therefore
${\rm BR}_\gamma$ should not be greater than about 0.01.  However,
using the typical values of parameters of interest for our present
application, we find ${\rm BR}_\gamma\sim 10^{-7}$, far below the 
sensitivity of current searches.

\subsection{Cosmic ray and CMB constraints}

DM annihilations can occur even after freezeout, with the production
of gamma rays or charged particles that can have an observable
effect. In the class of models we consider, the DM annihilates
directly into hidden sector gauge bosons, which in turn decay into
any charged SM particles that are sufficiently light.  Gamma rays
emerge  only as secondary products of these charged particles.  Their
contribution to the diffuse gamma ray background can potentially give
interesting constraints \cite{arina}, but currently the uncertainties
from details of structure formation do not allow one to draw firm
conclusions.  The production of antiprotons in the galaxy gives more
definite constraints, which can be quite stringent \cite{lavalle}.  
To avoid them, we need to assume that the gauge bosons which mix with
the photon are lighter than $2m_p$ so that $p\bar p$ pairs are not 
produced.  

\begin{figure}[t]
\smallskip \centerline{
\epsfxsize=0.45\textwidth\epsfbox{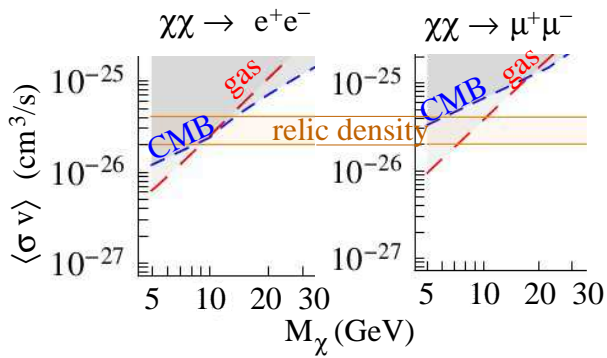}}
\caption{Adapted from ref.\ \cite{CMB}, showing constraints on 
the DM annihilation cross section versus mass from optical depth 
(``CMB'')
and excess heating of the intergalatic medium (``gas'').  Regions above
the diagonal lines are excluded. ``Relic density'' region indicates the
desired value of the cross section for the correct thermal abundance.}
\label{cmb}
\end{figure}

Charged particles that are produced around the time of recombination
reionize the plasma and  change the optical depth to the surface of
last scattering, a quantity that affects the Doppler peaks of the
cosmic microwave background \cite{cmb,CMB}. The effect is particularly
strong for DM with mass $M_\chi \lsim 10$ GeV, as in the exothermic
proposal for DAMA.  Ref.\ \cite{CMB} shows that such DM is marginally
ruled out if it decays exclusively into $e^+e^-$, while it is
marginally allowed if it decays into heavier charged particles (which
decay into electrons that are less energetic than if they were
primary products).  The relevant bounds are reproduced in fig.\ 
\ref{cmb}.   In most of our examples, the gauge bosons can decay into
muons and charged pions, so the branching ratio into electrons will
be suppressed and the bound should be somewhere between the two cases
shown in fig.\ \ref{cmb}.  Ref.\ \cite{cmb} also derives a bound from 
excess heating of the interstellar gas, which is more stringent than
the CMB bound in this small $M_\chi$ region, but which is also more
subject to uncertainty because of its dependence upon assumptions
about the details of structure formation.

\subsection{Long-lived gauge bosons and nucleosynthesis}

It is interesting to consider possible effects of the hidden
sector gauge bosons in the early universe.  Decays around the time
of big bang nucleosythesis or later can be deleterious, although
it is also possible to improve the predictions of big bang
nucleosynthesis (BBN), notably
for lithium \cite{posp-prad}.  In our models, $B_1$ is usually the
lightest gauge boson, and it couples to electrons (and muons and pions) with
$\epsilon_1\sim 10^{-5}$, fixed by the rate of direct DM detection.
Its decay rate is therefore of order $\alpha\epsilon_1^2 \mu_1$
which for $\mu_1\sim 500$ MeV leads to a lifetime of $10^{-10}$ s,
which is clearly harmless.

On the other hand, if $\epsilon_2\lsim 10^{-10}$ in the exothermic
model, where we have the constraint $\epsilon_2 \lsim 10^{-8}$, then
$B_2$ can have a lifetime greater than 1 s and possibly be relevant
for nucleosynthesis.  The question is whether its relic density is
large enough to have an effect.  We have computed the cross section
for $B_2 B_2\to B_1 B_1$ using FeynCalc \cite{feyncalc}.  
The cross section 
as $v\to 0$ can be expressed as
\beq
	\sigma v = {\pi\alpha_g^2\over 2\mu_2^2}f(x)
\eeq
where $x = 1-\mu_2/\mu_3= 2\delta M_{23}/\alpha_g \mu_3$ and
$f$ has a minimum value of 18 at $x=0$
(treating (\ref{mu1eq}) as an equality to eliminate $\mu_1$).
This neglects dark Higgs exchange in the $s$-channel, but we have 
checked that including it makes no dramatic  
change unless the virtual Higgs goes on shell.  For typical values
we find that the standard relic abundance calculation gives a
freezeout temperature around 8 MeV for $B_2$, and an abundance
$10^{-2}$ times smaller than that of the baryon asymmetry.  This is
too small to have any effect on BBN.

Ref.\ \cite{posp-prad} point out that a more likely candidate for
giving interesting effects is the dark Higgs bosons.  In particular,
if there exist a Higgs boson that is lighter than the gauge bosons,
it would decay into 4 leptons through emission of two virtual gauge
bosons, with a rate suppressed by $\epsilon^2\alpha_g\alpha^2
(m_h/\mu)^8$.  The annihilation cross section is suppressed for
similar reasons.  This can more naturally give long-lived 
relics (on the time scale of BBN) that could solve the lithium 
problem.

\section{Laboratory searches}
\label{lab}

\subsection{Beam dump experiments}

An interesting feature of the class of models we consider is that
they can be tested in proposed low-energy laboratory experiments.
A beam dump on an absorbing target can produce the weakly interacting
$B$ bosons that mix with the photon.  These can decay into $e^+e^-$
or other charged particles before reaching the detector, providing a
signal not present in the standard model.  

In our scenario, two of the three colors of bosons should mix with
the photon: $B_1$, with strength $\epsilon_1\sim 10^{-3}-10^{-5}$  to
get the right rate of direct detection, and either $B_3$ or $B_2$,
depending upon whether $\chi_2$ or $\chi_3$ is the stable excited
state.  Let us denote the corresponding mixing parameter by
$\epsilon_{3,2}$.  We noted in section \ref{kineq} (figure
\ref{kindec2}) that this parameter is essentially unconstrained.
If $\epsilon_{3,2}\gsim 10^{-6}$ then the effects of 
$B_{3,2}$ could be discovered in laboratory searches.  But since we
have more definite predictions for $B_1$, we will focus here on
its discovery potential.  Moreover we have argued that there are
certain advantages to having very small values of $\epsilon_{3,2}$
which could make laboratory detection of $B_{3,2}$ impossible for
the present.

\begin{figure}[t]
\smallskip \centerline{
\epsfxsize=0.45\textwidth\epsfbox{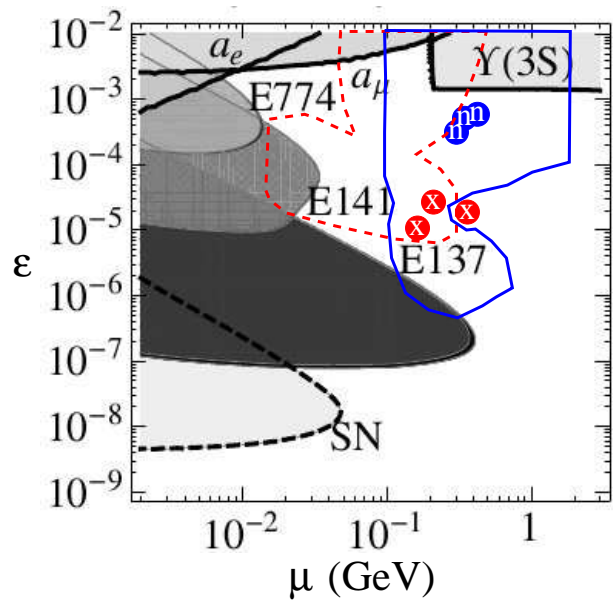}}
\caption{Potential for discovery of light  mixed
gauge bosons in plane of kinetic mixing parameter $\epsilon$ and
gauge boson mass $\mu$.  
Shaded regions are ruled out by existing laboratory or astrophysical
constraints.  Unshaded enclosed regions denote the reach of
experimental strategies proposed in ref.\ \cite{toro}.  Circles
containing ``n'' or ``x'' are typical predictions of our endothermic
or exothermic DM models, respectively.}
\label{bj}
\end{figure}

The authors of ref.\ \cite{toro} has recently summarized the current
experimental constraints in the $\epsilon_i$-$\mu_i$ plane (where
$\mu_i$ is the mass of the relevant gauge boson), and they have also
proposed strategies for new experiments that can cover more of the
still-allowed region in this plane.  Fig.\ \ref{bj} reproduces some
of their results.  On top of these we plot several examples of
predictions from our endothermic and exothermic models (circles
containing ``n'' or ``x'' respectively), corresponding to those
shown in figures \ref{endothermic}, \ref{exothermic} (see also 
fig.\ \ref{eps-mu}).  Almost all of these points are contained within
the contours denoting the reach of feasible new experiments suggested
by ref.\ \cite{toro}: the solid (blue) line denoting the high
resolution, high rate trident spectromenter, and the dashed (red) one
for the thin-target with double arm spectrometer.  It is suggested
that such experiments would be feasible at several existing 
laboratories, including Jlab (Thomas Jefferson National Accelerator
Facility), SLAC (Stanford Linear Accelerator Center), ELSA (Electron
Stretcher and Accelerator), and MAMI
(Mainzer Mikrotron).  We see that only
one of our examples (the right-most ``x'') would lie outside of the 
reach of the proposed experiments.  This corresponds to the extreme
case where $\mu_2 = 2$ GeV in fig.\ \ref{endothermic}.  The more
typical models would therefore be in the discoverable region.

\subsection{Invisible width of $Z$ boson}
The nonabelian gauge kinetic mixing portal (\ref{kinmix}) provides
two invisible decay channels for the $Z$ boson: $Z\to\Delta_i B_i$
since $\epsilon_i$ stands for the VEV of the Higgs triplet $\Delta_i$
over the heavy scale $\Lambda$ (eq.\ (\ref{lag})), and $Z\to B_j B_k$
where $i,j,k$ are a cyclic permutation of $1,2,3$.  The latter arises
because the nonabelian field strength $B_i^{\mu\nu}$ contains
$g\epsilon_{ijk}B_j^\mu B_k^\nu$.  Considering the first process, the
partial width is
\beq
	\Gamma_{Z\to\Delta B} = {m_Z^3\over 96\pi\,\Lambda^2}
\eeq
in the approximation $m_Z$ is much greater than the masses of the
decay products.  Demanding that this be less than the experimental
error on the invisible $Z$ width, $1.5$ MeV \cite{PDG}, we find
that $\Lambda > 1.3$ TeV.  For $\langle\Delta\rangle \sim 10$ GeV,
this leads to the bound $\epsilon \lsim 10^{-2}$, which is less
stringent than other constraints shown in fig.\ \ref{bj}.  For the
$Z\to B_j B_k$ channel, the partial width is of order
$\alpha_g\epsilon^2 m_z$.  This leads to a weaker bound on $\epsilon$
than does the $Z\to\Delta_i B_i$ channel.

\section{Conclusions}
\label{conc}

If the anomalous 511 keV gamma rays from the galactic center are
truly distributed in an axisymmetric manner, as suggested by the 
INTEGRAL observations, this provides strong motivation to seriously
consider DM decays or scatterings as their source, rather than 
localized sources such as supernovae or x-ray binaries.  A new
measurement with higher spatial resolution would be very desirable to
help settle this question. In the meantime, it seems worthwhile to
explore possible DM interpretations, especially if they can explain
more than just the 511 keV signal.  In the present work we have shown
how a three-component DM model with two mass splittings and a hidden
SU(2) gauge boson might address both the 511 keV observation and
indications of DM detection by DAMA/LIBRA and possibly CoGeNT.

The scenarios we have presented involve slightly inelastic nuclear
scatterings, either endothermic or exothermic, in the direct
detection experiments: $\chi_{2,3} N \to \chi_{3,2} N'$.  The
endothermic version with $M_\chi\cong 4$ GeV is under stronger
pressure from astrophysical constraints from the CMB (figure
\ref{cmb}) and especially from the decay $\chi_3\to\chi_2\gamma$, not
observed by Chandra (section \ref{spdms}).  The latter could possibly
be softened by some modification of the particle physics model, such
as extending the gauge group to SU(2)$\times$U(1). The exothermic
model also requires a more cuspy halo than  the endothermic one to
get the observed 511 keV rate from  $\chi_3\chi_3\to\chi_2\chi_2$
scattering, although still consistent with examples from $N$-body
simulations that take into account compression by baryons.

We have highlighted two distinct mechanisms for getting the 511
keV signal: either inelastic $\chi_{2,3}\chi_{2,3}\to
\chi_{3,2}\chi_{3,2}$ scatterings followed by $\chi_{3,2}\to \chi_1
e^+ e^-$ decays, or the decay process by itself when  $\chi_{3,2}$
has a lifetime of order $10^5$ times the age of the universe. 
Whereas the first mechanism requires some mutual adjustments of the
particle physics and DM halo parameters to get the right rate, the
second is more easily arranged by just tuning the gauge kinetic
mixing parameter $\epsilon_{2,3}\sim 10^{-11}$ that controls the
decay rate.  The decay mechanism points to the exciting
possibility that the angular profile of the 511 keV signal is
actually a picture of  the DM halo profile in the inner part of the
galaxy, if positron diffusion is a negligible effect.  The scattering
mechanism on the other hand {\it requires} significant positron
diffusion, or else propagation of the excited DM state before decay,
since otherwise it predicts too narrow angular profile.  It is
interesting that our model can naturally explain such
long-distance travel of the excited DM prior to its decay, by tuning
$\epsilon_{2,3}\sim 10^{-7}$.  

A very encouraging aspect of these proposals is their testability in
low-energy electron beam dump experiments.  The  kinetic mixing
parameter $\epsilon_1\sim 10^{-4}$ and  the mass $\mu_1 \lsim 1$ GeV 
of the gauge boson mediating  the direct detection scatterings are in
prime territory for their discovery by such experiments, which could
be done at existing laboratories.   The models presented here are
also potentially rich in consequences for cosmic rays, the diffuse
x-ray or gamma-ray backgrounds, the CMB, and big bang
nucleosynthesis.

\bigskip

{{\bf Acknowledgment}.  We thank Celine Boehm, Laurent Bouchet,
Marco Cirelli, Neal
Dalal, Malcolm Fairbairn, Jonathan Feng, Katia Ferriere, Ben
Grinstein, Pierre Jean, Manoj Kaplinghat,  Vicky Kaspi,  Jonathan
McDowell, Guy Moore, Nicolas Produit, Gerald Skinner, 
and Tracy Slatyer for helpful
correspondence  or discussions.  Our work is supported by the Natural
Sciences and Engineering Research Council (NSERC) of Canada.  JC
thanks the CERN theory group and the Perimeter Institute
for their hospitality during part of this work.}

\appendix
\section{Annihilation amplitudes and rates}\label{thermalrelic}

In this appendix, we derive the invariant amplitudes (squared) for
annihilation of DM particles
$\chi$ to both gauge and Higgs bosons, including the lowest order 
corrections due to dark matter velocity, which we use to find the relation
between the thermal relic density and the dark gauge coupling $\alpha_g$
in section \ref{cross_alphag}.  Including annihilation to Higgs bosons 
extends and corrects the results listed in \cite{nonabelian}; in addition, we
correct the final state polarization and color sums carried out in
\cite{nonabelian}.
To keep the final result simple, we will first assume that
symmetry breaking occurs at a lower temperature than DM freezeout, so the
gauge and Higgs bosons may be treated as massless.  The effects of symmetry
breaking are discussed at the end.  For reference, we will 
consider a general gauge group and general representations for both DM and the
Higgs.

We consider first DM annihilation to Higgs bosons.  This process is mediated
by t- and u-channel diagrams involving Yukawa couplings at each vertex,
s-channel diagrams with an intermediate Higgs particle connecting 
a Yukawa coupling at one vertex  to
a scalar potential vertex, and an s-channel diagram with an intermediate
gauge boson coupling to the DM and Higgs particles at either end.
Since we are concerned in
this paper with either parametrically small or vanishing Yukawa couplings,
we assume that the s-channel diagram with an intermediate gauge boson 
dominates.  

Consider incoming DM states $\chi_i$ in representation $R$ and outgoing Higgs
states $\Delta_I$ in representation $R'$; the incoming momenta are $p_{i}$ and 
outgoing momenta $q_I$.  The matrix element is\footnote{We use a mostly
plus metric with Dirac algebra $\{\gamma^\mu,\gamma^\nu\}=-2\eta^{\mu\nu}$.} 
\beq \mathcal{M} = \frac{ig^2}{(p_i+p_j)^2} \bar v_j T^a_{ji} \gamma^\mu u_i 
T^a_{JI}(q_I-q_J)_\mu\ .\eeq
Once summed over outgoing
colors and averaged over incoming colors and spins, it is
\beqa
|\mathcal{M}|^2 &=& \frac{1}{4d_R^2} \frac{g^4}{s^2} 
\mathrm{tr}\left[ (\slashed{p_i}+M_\chi)
(\slashed{q}_I-\slashed{q}_J)
(\slashed{p}_j-M_\chi)(\slashed{q}_I-\slashed{q}_J)\right]\nonumber\\
&\times&T^b_{ji} T^a_{ij}T^b_{JI} T^a_{IJ}\ ,
\eeqa
where $d_R$ is the dimension of representation $R$.  The color sums both
take the form 
\beq \mathrm{tr}_R T^b T^a = (d_R/d_{adj}) C_2(R) \delta^{ab}\ ,\eeq
where $C_2$ is the quadratic Casimir; the sum over the adjoint indices
gives $\delta^{ab}\delta^{ab}=d_{adj}$.  For nonrelativistic dark matter
at center-of-mass velocity $v=v_{\rm rel}/2$ and scattering angle $\theta$, 
the amplitude becomes
\beq |\mathcal{M}|^2 = \left(\frac{d_{R'}}{d_{adj} d} C_2(R) C_2(R')\right)
\frac{g^4}{2}\left(1-v^2\cos^2\theta\right)\ .\eeq

Annihilation to gauge bosons (of colors $a,b$ and momenta $q_{a,b}$) 
receives contributions from s-, t-, and u-channels.  The amplitudes for
each channel are (as in \cite{nonabelian}) 
\beqa
\mathcal{M}_s &=&\!\!\! \frac{g^2}{s} \bar v_j T^c_{ji} \gamma_\lambda u_i
f^{abc}\varepsilon^\star_\mu(a)\varepsilon^\star_\nu(b)
\left[\eta^{\mu\nu}(q_b-q_a)^\lambda\right.\nonumber\\
&&\left.-\eta^{\nu\lambda}(q_b+p_i+p_j)^\mu
+\eta^{\mu\lambda}(q_a+p_i+p_j)^\nu\right],\nonumber\\
\mathcal{M}_t &=&\!\!\! i\frac{g^2}{t-M_\chi^2} T^b_{jk} T^a_{ki}\bar v_j 
\slashed{\varepsilon}^\star(b) (\slashed{p}_i-\slashed{q}_a-M_\chi)
\slashed{\varepsilon}^\star(a)u_i\ ,\nonumber\\
\mathcal{M}_u &=&\!\!\! i\frac{g^2}{u-M_\chi^2} T^a_{jk} T^b_{ki}\bar v_j 
\slashed{\varepsilon}^\star(a) (\slashed{p}_i\!-\!\slashed{q}_b\!-\!M_\chi)
\slashed{\varepsilon}^\star(b)u_i\! .
\eeqa
We need to account for both direct squares and cross terms in the amplitudes.
After some tedious algebra including Dirac traces, we find the following 
results for the (color and spin summed and averaged) square amplitude:
\beqa
|\mathcal{M}_s|^2 &=& \frac{g^4}{d_R^2} \mathrm{tr}_R T^c T^d f^{abc} f^{abd}
\left(-\frac{19}{4}+\frac{1}{8}v^2\left(11\right.\right.\nonumber\\
&&\left.\left.-5\cos2\theta\right)\vphantom{\frac{1}{8}}\right)\ ,\nonumber\\
|\mathcal{M}_{t,u}|^2 &=&\frac{2g^4}{d_R^2}
\mathrm{tr}_R \left(T^a T^b T^b T^a\right) \left(1\pm v\cos\theta +v^2\right)
\ ,\nonumber\\
\mathcal{M}_s\bar{\mathcal{M}}_{t}&=&\frac{ig^4}{d_R^2} f^{abc}\mathrm{tr}_R
\left(T^cT^bT^a\right)\left(2-v^2\sin^2\theta\right)\ ,\nonumber\\
\mathcal{M}_s\bar{\mathcal{M}}_{u}&=&-\frac{ig^4}{d_R^2} f^{abc}\mathrm{tr}_R
\left(T^cT^aT^b\right)\left(2-v^2\sin^2\theta\right)\ ,\nonumber\\
\mathcal{M}_t\bar{\mathcal{M}}_{u}&=&\frac{2g^4}{d_R^2} \mathrm{tr}_R
\left(T^a T^b T^a T^b\right) v^2\ .
\eeqa
Here, we have expanded around zero DM velocity as for annihilation to 
Higgs bosons.

We can evaluate the group theory factors using 
$f^{abc}f^{abd}=C_2(adj)\delta^{cd}$ and $T^a T^a=C_2(R)$, antisymmetry of
the structure constants, and the group algebra.  We find
\beqa
|\mathcal{M}_s|^2 &=& \frac{g^4}{d_R}C_2(adj)C_2(R)
\left(-\frac{19}{4}+\frac{1}{8}v^2\left(11\right.\right.\nonumber\\
&&\left.\left.-5\cos2\theta\right)\vphantom{\frac{1}{8}}\right)\ ,\nonumber\\
|\mathcal{M}_{t,u}|^2 &=&\frac{2g^4}{d_R}C_2(R)^2 
\left(1\pm v\cos\theta +v^2\right)
\ ,\nonumber\\
\mathcal{M}_s\bar{\mathcal{M}}_{t}&=&\frac{g^4}{2d_R} C_2(adj)C_2(R)
\left(2-v^2\sin^2\theta\right)\ ,\nonumber\\
\mathcal{M}_s\bar{\mathcal{M}}_{u}&=&\frac{g^4}{2d_R} C_2(adj)C_2(R)
\left(2-v^2\sin^2\theta\right)\ ,\nonumber\\
\mathcal{M}_t\bar{\mathcal{M}}_{u}&=&\frac{2g^4}{d_R} \left(C_2(R)^2
+\frac{i}{2}C_2(adj)C_2(R)\right)v^2\! .
\eeqa

In the above discussion, we took the sum over gauge boson polarization
vectors to give the metric for simplicity; this includes longitudinal and
timelike polarizations as well as the transverse ones.  
In a nonabelian gauge theory, the unphysical polarizations do not 
automatically vanish when contracted in the amplitudes, so we must correct
for their inclusion.  We can do this by \textit{subtracting} the 
squared amplitude for ghost production.  The amplitude is
\beq
\mathcal{M}=\frac{g^2}{s}\bar v_jT^c_{ji}\slashed{q}_a u_i f^{abc}\ ,
\eeq
so we find
\beq
|\mathcal{M}|^2=\frac{g^4}{8d_R} C_2(adj)C_2(R)\left(1+v^2(1-\cos^2\theta)
\right)\ .\eeq
As an example of this effect, we can consider $\chi_1\chi_2\to B_1 B_2$ 
scattering (with fixed colors for $SU(2)$ triplet DM).  
Taking just the kinematical factors,
the amplitude for annihilation into gauge bosons is given by
$|\mathcal{M}_s+\mathcal{M}_u|^2=5g^4/4$, as in \cite{nonabelian}.  The 
ghosts subtract $g^4/8$ for a total of $|\mathcal{M}|^2=9g^4/8$, in 
agreement with the massless limit of the amplitude in the symmetry breaking
phase.

We are primarily interested in $SU(2)$ triplet DM with triplet and quintuplet 
Higgs fields, and we now specialize to that case, assuming $N_3$ Higgs 
fields in the triplet and $N_5$ in the quintuplet.  
The total squared amplitude, with gauge and Higgs boson final states added
incoherently, is 
\beqa 
|\mathcal{M}|^2 &=& \frac{g^4}{3}\left[\left(\frac{25}{2}+2N_3 
+10N_5 \right)+\frac{63}{2}v^2\right.\nonumber\\
&&\left. +\left(\frac{7}{2}-2N_3-10N_5\right)v^2\cos^2\theta\right]\ .\eeqa
For identical nonrelativistic initial particles and identical massless final
particles, the differential cross-section is
\beq \frac{d\sigma}{d\Omega} = \frac{1}{2} \frac{1}{64\pi^2 s}
\frac{|\vec q_a|}{|\vec p_i|} |\mathcal{M}|^2\ ,\label{diffcross}\eeq
where the factor of $1/2$ is due to overcounting identical final states 
in the color sum.  (Here we have assumed that the Higgs particles are in a
real representation; if not, drop the factor of $1/2$ for the Higgs final
states.) 

So far we have worked at tree-level and in the CM frame, but there are small
corrections to both approximations.  First, the annihilation cross section
experiences Sommerfeld enhancement since the DM is nonrelativistic.  
Under the assumption that the gauge symmetry is not yet broken, the boost
factor is 
\beq S(v) = \frac{\pi \alpha_g/v}{1-e^{-\pi\alpha_g/2v}}\ ,\ \ 
\alpha_g =\frac{g^2}{4\pi}\ .\eeq
In the parameter space appropriate to the thermal DM freezeout, 
$\pi\alpha_g/v<1$, so we treat it as a small parameter and expand
$S(v)\sim 1+\pi\alpha_g/2v$.  Next, since the CM frame is not quite the
rest frame of the cosmic fluid, we must include the Lorentz transformation
of the cross section. For nonrelativistic center-of-mass velocity $\vec V$, 
this correction
takes $\sigma\to \sigma (1-V^2\sin^2\phi/2)$, where $\phi$ is the angle
between $\vec V$ and $\vec v$.  

Finally, we must average the cross section over the DM velocity distribution,
which is Maxwell-Boltzmann.  In terms of the $\vec V$ and $\vec v$, this
average takes the form
\begin{widetext}
\beqa \langle\sigma v_{\rm rel}\rangle&=& \left(\frac{M_\chi}{2\pi T}\right)^3
\int d^3\vec V d^3\vec v\, (\sigma v_{\rm rel})(v)\left(1-\frac{1}{2}V^2\sin^2\phi
\right) e^{-M_\chi(v^2+V^2)/T}\nonumber\\
&=&\frac{\pi}{12}\frac{\alpha_g^2}{M_\chi^2} \left[\left(\frac{25}{2}+2N_3+10N_5
\right)\left(1+\alpha_g\sqrt{\frac{\pi M_\chi}{T}}
-\frac{1}{2\pi}\frac{T}{M_\chi}\right)
+\left(\frac{317}{8}-\frac{5}{2} N_3-\frac{25}{2}N_5
\right)\frac{T}{M_\chi}\right]
\ ,\eeqa
\end{widetext}
where the relative velocity $v_{\rm rel}=2v$.  We have expanded the result to 
first order in the small parameters $\alpha_g\sqrt{M_\chi/T}$ and $T/M_\chi$.  
In the sequel, we will treat the ratio $M_\chi/T$ as roughly independent
of the DM mass and use an approximate value of $M_\chi/T\sim 20$.

We should also address the issue of symmetry breaking.  We will assume that
the gauge bosons are light compared to the DM and that Higgs bosons are 
either light compared to the DM or too heavy to be produced in DM annihilation.
The annihilation amplitudes are slightly modified by the gauge boson masses
(in a manner that does not respect the gauge symmetry, of course).  
For example, $\chi_1\chi_2\to B_1B_2$ annihilation has 
\beq |\mathcal{M}|^2= g^4\frac{\mu_1^4+2\mu_1^2(5\mu_2^2-\mu_3^2)+
(\mu_2^2-\mu_3^2)^2}{8\mu_1^2\mu_2^2}\eeq
at zero velocity and lowest order in the gauge boson masses.  This deviates
from the leading order massless result only in as much as the gauge boson
masses differ from each other.  Note that 
the Goldstone boson states become longitudinal gauge bosons in gauges such
as the unitary gauge.  The significant effect of symmetry breaking
is to change the kinematical factors in the cross section.  Using the
average mass $\bar\mu$ for all the light gauge bosons and Higgs states,
\beqa |\vec q_a|&=&M_\chi(1+v^2/2) \to \sqrt{M_\chi^2(1+v^2)-\bar\mu^2}
\nonumber\\
&\cong&\!\! M_\chi\!\left(1-\frac{\bar\mu^2}{M_\chi^2}\right)^{1/2}\!\!\left(1+
\frac{M_\chi^2v^2}{2(M_\chi^2-\bar\mu^2)}\right)\!\! .\eeqa
For symmetry breaking at a small scale compared to the DM mass, this just
renormalizes the cross section by a factor of $(1-\bar\mu^2/M_\chi^2)^{1/2}$.

We can now compare the cross section to that required for the correct relic
density of dark matter.  Under the normal assumption that only SM particles
are lighter than the dark matter, the required cross section is approximately
$\langle\sigma v_{\rm rel}\rangle_0\sim 2.84\times 10^{-26}\ 
\textnormal{cm}^3/\textnormal{s}$, with a logarithmic dependence on DM mass.
We will take this central value.  However, our dark matter models contain
extra light degrees of freedom, which affects the required cross section in
two ways.  First, the Hubble parameter is greater at a given temperature,
which causes freezeout to occur earlier.  Second, due to heating of photons
by annihilation of these light degrees of freedom (see Big Bang 
nucleosynthesis constraints given in \cite{nonabelian}), 
the universe expands more
between freezeout and the present day, which means that the freezeout 
density of DM must be higher than in minimal DM models.  As a result,
the desired cross section satisfies
\beq \langle\sigma v_{\rm rel}\rangle = 
\frac{\sqrt{g_*}/g_{*s}}{(\sqrt{g_*}/g_{*s})_0}
\langle\sigma v_{rel}\rangle_0\ .\eeq
For DM masses near $M_\chi=5$ GeV, this is
\beq \langle\sigma v_{\rm rel}\rangle=
\frac{\langle\sigma v_{\rm rel}\rangle_0}{\sqrt{1+(6+3N_3+5N_5)/61.75}}\ .\eeq

Including all these corrections, we can write the desired cross section
in terms of the gauge coupling as
\beq \langle\sigma v_{\rm rel}\rangle = (A\alpha_g^2+B\alpha_g^3)/M_\chi^2\ ,
\eeq
which has the iterative solution
\beq
\alpha_g\cong \left(\frac{\langle\sigma v_{\rm rel}\rangle}{A+B\alpha_g}
\right)^{1/2} M_\chi\ .\eeq
Our $SU(2)$ models have 
\beqa \alpha_g&\cong& \frac{M_\chi}{\textnormal{GeV}}
\left(1-\frac{\bar\mu^2}{M_\chi^2}\right)^{-1/4}\nonumber\\
&\times&
\left\{\begin{array}{cc} 2.5\times 10^{-5}& (N_3=0, N_5=0;\ 
\textnormal{all } m_h>M_\chi)\\
2.2\times 10^{-5}  & 
(N_3=2, N_5=0)\\ 2.0\times 10^{-5} & (N_3=3, N_5=0)\\
1.7\times 10^{-5} & (N_3=2, N_5=1) \\ 1.7\times 10^{-5} & (N_3=3, N_5=1)
\end{array}\right.\ .\label{alphagfinal}
\eeqa
In the end, corrections due to the initial velocity of the DM particles 
contribute at the $5-15\%$ level, while corrections from Sommerfeld 
enhancment contribute only 1 part in $10^4$ due to their additional
dependence on $\alpha_g$ (since the coefficient $B$ is of the same order
as $A$).  

\section{Kinetic coupling to SM}\label{kineticSM}
In this appendix, we find the freezeout temperature of the
dark matter kinetic coupling to the Standard Model.
For convenience calculating phase space factors, 
we consider downscattering $\chi_{2,3} f\to \chi_1 f$, labeling the
$\chi_j$ momenta as $p_j$ and the initial and final energy and momenta of the 
SM fermions $f$ as $E_{i,f}$ and 
$q_{i,f}$ respectively (similarly for other subscripts).  Here, the
relevant initial DM state is $\chi_3$ for the endothermic case and $\chi_2$
for exothermic.  In this appendix, we consider the endothermic case, but
it should be clear that these results apply equally well in both scenarios.
At temperatures under 100 MeV (and, in particular near the important scale
of $\delta M_{13}\sim$ MeV), only scattering from $e^{\pm}$
will be important.  Furthermore, if this process occurs roughly once per
Hubble time for each of the more massive DM particles, it efficiently
maintains the distribution of these two DM states given by the Boltzmann
factor at the SM temperature.
Also, if $\chi\chi$ scattering is still in 
equilibrium, this reaction can insure that all the DM states maintain a
thermal velocity distribution at the SM temperature.  
(As we have seen in the main text, 
$\chi\chi$ scattering typically freezes out later than this process.)
We note that the same calculations apply for scattering between the
two top states $\chi_3 f\to \chi_2 f$ with the appropriate replacements of
$\epsilon_2$, $\mu_2$, and $\delta M_{13}$.  As shown in figure \ref{kindec},
the cross section is only slightly smaller for keV mass splittings.

The matrix element for the scattering process shown in figure \ref{scatt}
is (taking electrons for specificity)
\beq
\mathcal{M} = -i \frac{ge\epsilon_2}{(p_3-p_1)^2-\mu_2^2}\bar u_1\gamma^\mu u_3
\bar u_f\gamma_\mu u_i\ .
\eeq
After the spin sum and average,
\begin{widetext}
\beqa
\frac{1}{4}\sum |\mathcal{M}|^2 &=& 4\frac{(ge\epsilon_2)^2}{(\mu_2^2 -t)^2}
\left[s^2+\frac{1}{2}t^2+st
-s(2m_e^2+2M_\chi^2 +2M_\chi\delta M_{13}+\delta M_{13}^2)
-\frac{1}{2} t\delta M_{13}^2\right.\nonumber\\
&&\left. +M_\chi^4+2 M_\chi^3\delta M_{13} +M_\chi^2\left(\delta M_{13}^2
+2m_e^2\right)
+2M_\chi\delta M_{13}m_e^2 +m_e^4\vphantom{\frac{1}{2}}
\right]\ .\label{ampsquared}\eeqa
\end{widetext}
With the replacement $m_e\to m_p$, $\delta M_{13}\to\delta M_{23}$, and
$\mu_2\to\mu_1$,
this is also the result for inelastic DM scattering off protons, 
which is relevant to direct detection experiments.

Since we are interested in temperatures much less than the DM mass $M_\chi$
(and all other energy scales are also much less than $M_\chi$),
we can work in center-of-momentum (CM) frame up to overall errors of order
$\sqrt{T/M_\chi}\ll 1$ in the cross-section
compared to the cosmic rest frame.  To lowest order
in $T/M_\chi$, the final electron energy is
\beq E_f = E_i+\delta M_{13}\ ,\eeq
or
\beq |\vec q_f|^2 = \delta M_{13}^2+2\delta M_{13} E_i
+|\vec q_i|^2\ .\eeq
Then the Mandelstam $t$ ranges between $t_-$ and $t_+$ satisfying 
\beqa
t_\pm &=& -2|\vec q_i|^2 -2\delta M_{13} E_i
\nonumber\\
&&\pm 2|\vec q_i|\left[\delta M_{13}^2+2\delta M_{13}E_i
+|\vec q_i|^2\right]^{1/2} \eeqa
to lowest order.  Due to cancellations in (\ref{ampsquared}), we will
need $s$ to the same (second) order in small quantities:
\beq s = M_\chi^2 +2M_\chi\left(\delta M_{13} +E_i\right)+\delta M_{13}^2
+2\delta M_{13} E_i+m_e^2+2|\vec q_i|^2\ .
\eeq
To lowest order in small quantities, we find
\beq
|\mathcal{M}|^2 = 16\frac{(ge\epsilon_2)^2 M_\chi^2}{(\mu_2^2-t)^2}
\left[E_i^2+\delta M_{13}E_i+\frac{1}{4} t\right]\label{ampsimple}\eeq
for the (spin averaged and summed) squared amplitude.
Since $d\sigma/dt = (1/64\pi s)|\mathcal M|^2/|\vec q_i|^2$ and the relative
velocity is dominated by the electron velocity, we find that
\begin{widetext}\beq
\sigma v_{\rm rel} = \frac{(ge\epsilon_2)^2}{4\pi} \frac{1}{E_i|\vec q_i|}
\left[\left(E_i^2+\delta M_{13}E_i+\frac{\mu_2^2}{4}\right)
\left(\frac{1}{\mu_2^2-t_+}
-\frac{1}{\mu_2^2-t_-}\right)-\frac{1}{4}\ln\left(
\frac{\mu_2^2-t_-}{\mu_2^2-t_+}\right)\right]\ .\label{kinxsect}
\eeq
\end{widetext}



We are especially interested in whether the DM and SM can maintain kinetic
equilibrium at temperatures $T\lesssim m_e$, since those temperatures are
relevant for $\chi -\chi$ downscattering.  Then temperatures
are parametrically less than $\mu_2$, so
\beqa
\sigma v_{\rm rel} &=& \frac{(ge\epsilon_2)^2}{2\pi}\frac{1}{\mu_2^4}
\frac{1}{E_i}
\left(2m_e^2+|\vec q_i|^2 +\delta M_{13} E_i\right)\nonumber\\
&\times& \left(\delta M_{13}^2
+|\vec q_i|^2+2\delta M_{13} E_i\right)^{1/2}
\eeqa
times corrections of relative order $t/\mu_2^2$ and $\sqrt{|\vec q_i|/M_\chi}$,
where $v_{\rm rel}$ is the CM frame electron speed.
At a fixed temperature $T$, we find the total scattering rate
by integrating over the Fermi-Dirac distribution for marginally relativistic
electrons (since $T\sim m_e$).  Assuming a thermal origin for the dark 
matter abundance so $\alpha_g\propto M_\chi$, 
this total rate can be written in terms of a normalized rate $\hat\Gamma$ as
\beq \langle n_e \sigma v_{\rm rel}\rangle \equiv \left(
\frac{\epsilon_2^2 M_\chi}{\mu_2^4}
\right)\hat\Gamma(\delta M_{13}, T)\ .\eeq
In $n_e$, we include both spin states of electrons and positrons.

Therefore, the ratio $3H/\hat\Gamma$ considered as a function of $T$ 
inverts to give the DM/SM decoupling temperature as a function of
$\epsilon_2^2 M_\chi/\mu_2^4$.  This decoupling temperature is shown in figure 
\ref{kindec} for several values of $\delta M_{13}$. 
Since we consider temperatures near the electron mass, we calculate the
effective species number $g_*$ numerically.  This includes heating of 
photons due to $e^\pm$ annihilation and also the neutrino density.  

Since the kinetic mixing is between the $B_\mu^a$ bosons and 
SM hypercharge and therefore includes mixing with $Z_\mu$, 
we can ask if the DM comes into equilibrium with neutrinos through the weak
force.  Above the electron mass, scattering from electrons will always 
dominate scattering from neutrinos because $m_Z\gg \mu_2$ (and because the
$B_\mu$ coupling to the weak current is suppressed by $\sim \mu_2^2/m_Z^2$)
\cite{nonabelian}.
However, below the electron mass, there are many more neutrinos than 
electrons, so neutrino scattering is potentially important.  Scattering from
neutrinos progresses through two Feynman diagrams similar to figure 
\ref{scatt}, one with a $B$ propagator and one with a $Z$ propagator, but there
is a relative sign between the two in the kinetic mixing.  Therefore, to
include both diagrams properly, we should replace 
\beq \frac{1}{(\mu_2^2-t)^2}\to \left[\frac{\mu_2^2/m_z^2}{\mu_2^2-t}
-\frac{1}{m_Z^2-t}\right]^2\sim \frac{t^2}{\mu_2^4 m_Z^4}\eeq 
in equation (\ref{ampsimple}) at low temperatures.  Also, taking
$\epsilon$ to be the $B$ coupling to the electric current, we should replace
$\epsilon\to \epsilon \tan\theta_w$.  Due to the fact that the two Feynman
diagrams nearly cancel, the cross-section is highly suppressed.  A 
straightforward estimate of the total scattering rate and comparison to the
Hubble rate indicates that $\epsilon$ would need to be of order $10^5$ 
for $\chi - \nu$ scattering to equilibrate at temperatures below $m_e$!

\section{$\chi_2\to\chi_1 e^+ e^-$ at one loop}\label{loopdecay}

In this appendix, we give a careful derivation of the lifetime for 
$\chi_2\to\chi_1 e^+ e^-$ decay at one loop level when $\epsilon_3=0$.
This is the dominant decay process we have been able to find for this case.

Consider the loops in figure \ref{chidecay}.  As noted above the figure,
the two diagrams nearly cancel due to the opposite signs of the nonabelian
3-gauge-boson couplings; in fact, they do not cancel completely only because
the two gauge bosons in the loop have different masses.  The complete 
amplitude can be written as
\beq
\mathcal{M}=ig 
\frac{\bar u_1\gamma^\mu u_2 \bar u_e L_\mu v_e}{q^2-\mu_3^2}\ ,
\eeq
where the spinors of the $\chi$ particles are labeled by their color (as
their momenta $k_{2,1}$ will be), we take the outgoing momenta of the
positron and electron to be $p_\pm$ respectively, and $q=k_2-k_1=p_+ +p_-$.

The momenta running counter-clockwise around the loops are $l+\delta p$ on the
electron line, $l-\bar p$ on the upper gauge boson line, and $l+\bar p$
on the lower gauge line, where $\bar p=q/2$ and 
$\delta p=(p_+ - p_-)/2$.  With these conventions, the loop integrals are
\begin{widetext}
\beqa L_\mu &=&i g e^2 \epsilon_1\epsilon_2 \int \frac{d^4l}{(2\pi)^4}
\left[\frac{\gamma^\nu(\slashed{l}+\delta\slashed{p}+m_e)
\gamma^\lambda}{(l+\delta p)^2-m_e^2}\right]
\left[\eta_{\mu\nu}(3\bar{p}-l)_\lambda +\eta_{\nu\lambda}(2l)_\mu
-\eta_{\mu\lambda}(l+3\bar p)_\nu\vphantom{\frac{}{}}\right]\nonumber\\
&\times&
\left[\frac{1}{[(l+\bar p)^2-\mu_2^2][(l-\bar p)^2-\mu_1^2]}
-(\mu_1\leftrightarrow\mu_2)\right]\ .\label{loop1}
\eeqa
As usual, we can rewrite the denominators with Feynman parameters as 
\beqa 
\int \frac{dxdydz\, 2\delta(1-x-y-z)}{
[l^2+2l\cdot(\bar p(y-x)+\delta p z)+\bar p^2(x+y)+\delta p^2 z-\mu_1^2 x
-\mu_2^2 y-m_e^2 z]^3}\nonumber\\
= 
\int
\frac{dxdydz\, 2\delta(1-x-y-z)}{[l^{\prime 2} +\delta p^2 z(1-z)
+\bar p^2 (x(1-x)+y(1-y)+2xy)-\mu_1^2 x -\mu_2^2 y-m_e^2 z]^3}\ ,
\label{feynmanparams}\eeqa
\end{widetext}
where $l'=l+\bar p(y-x)+\delta p z$ and we have used $\bar p\cdot\delta p=0$.
From the second form, it is clear that taking $\mu_1\leftrightarrow\mu_2$ is
the same as swapping the Feynman parameters $x\leftrightarrow y$.
Therefore, the only terms that survive taking the difference in (\ref{loop1})
must be antisymmetric in $x\leftrightarrow y$, and these must come from
shifting $l$ to $l'$.

In the end, we find
\begin{widetext}
\beqa L_\mu &=& -8 \left(\bar p_\mu \delta\slashed{p} 
+\delta p_\mu \bar{\slashed{p}}\right) \int_0^1 dx\int_0^{1-x} dy\int
\frac{d^4l}{(2\pi)^4} \frac{(y-x)(1-x-y)}{[l^2-\Delta(x,y)]^3}\nonumber\\
&\cong& \frac{-i}{12\pi^2} \left(\bar p_\mu \delta\slashed{p} 
+\delta p_\mu \bar{\slashed{p}}\right)\left[\frac{1}{\mu_2^2-\mu_1^2}+
\frac{\mu_1^2+\mu_2^2}{[\mu_2^2-\mu_1^2]^2}\ln\left(\frac{\mu_1}{\mu_2}
\right)\right]\ ,\label{loop2}
\eeqa
\end{widetext}
with $\Delta(x,y)$ given as in (\ref{feynmanparams}).  Note that this
loop appears to generate an interaction with one extra derivative compared
to a magnetic moment operator.  In the approximation,
we have taken the gauge boson masses to be much larger than any of the 
momenta in the denominator.  We also have $\bar u_e\delta\slashed{p}v_e=2m_e$
and $\bar u_e\bar{\slashed{p}}v_e=0$, so the amplitude is finally
\beq \mathcal{M}\cong -\frac{8\alpha_g\alpha\epsilon_1\epsilon_2 m}{3\mu_3^2}
\bar u_1\delta\slashed{p}u_2 \bar u_e v_e\Lambda\ ,\eeq
where $\Lambda$ is the function of $\mu_{1,2}$ in square brackets in 
(\ref{loop2}).

In the nonrelativistic limit, the spin-summed and averaged square amplitude
is
\beqa |\mathcal{M}|^2 &\cong& \frac{1024}{9}
\frac{\alpha_g^2\alpha^2\epsilon_1^2\epsilon_2^2 m_e^2 M_\chi^2}{\mu_3^4} 
\Lambda^2 E_+ E_-\nonumber\\
&\times&\left(E_+ E_- - \vec p_+ \cdot \vec p_- -m_e^2\right)\ ,\eeqa
with $E_\pm$ and $\vec p_\pm$ the energy and 3-momentum of the $e^\pm$.
Integrating over the nonrelativistic phase space, we find
\beq \Gamma = \frac{16}{9\pi^2} \alpha^2\alpha_g^2\epsilon_1^2\epsilon_2^2
\frac{m_e^5}{\mu_3^4} \left(\delta M_{12}-2m_e\right)^3
\left(\delta M_{12}-m_e\right)\Lambda^2\ .\eeq
Using the same estimated parameters as below equation (\ref{loopdecay1}),
we find a lifetime of more than $10^{33}$ s.


\begin{thebibliography}{99}


\bibitem{Diehl}
  R.~Diehl and M.~Leising,
  arXiv:0906.1503 [astro-ph.HE].

\bibitem{spi}

  J.~Kn\"odlseder {\it et al.},
  Astron.\ Astrophys.\  {\bf 411}, L457 (2003)
  [arXiv:astro-ph/0309442];

P.~Jean {\it et al.},
  Astron.\ Astrophys.\  {\bf 407}, L55 (2003)
  [arXiv:astro-ph/0309484].

 J.~Kn\"odlseder {\it et al.},
  Astron.\ Astrophys.\  {\bf 441}, 513 (2005)
  [arXiv:astro-ph/0506026];

  L.~Bouchet, E.~Jourdain, J.~P.~Roques, A.~Strong, R.~Diehl, F.~Lebrun and R.~Terrier,
  arXiv:0801.2086 [astro-ph].


\bibitem{astro}

  W.~Wang, C.~S.~J.~Pun and K.~S.~Cheng,
  Astron.\ Astrophys.\  {\bf 446}, 943 (2006)
  [arXiv:astro-ph/0509760];

M.~Casse, B.~Cordier, J.~Paul and S.~Schanne,
  Astrophys.\ J.\  {\bf 602}, L17 (2004)
  [arXiv:astro-ph/0309824].

 G.~Bertone, A.~Kusenko, S.~Palomares-Ruiz, S.~Pascoli and D.~Semikoz,
  Phys.\ Lett.\  B {\bf 636}, 20 (2006)
  [arXiv:astro-ph/0405005].

  P.~A.~Milne, J.~D.~Kurfess, R.~L.~Kinzer and M.~D.~Leising,
  New Astron.\ Rev.\  {\bf 46}, 553 (2002)
  [arXiv:astro-ph/0110442].

 K.~Ahn, E.~Komatsu and P.~Hoflich,
  Phys.\ Rev.\  D {\bf 71}, 121301 (2005)
  [arXiv:astro-ph/0506126].


  R.~M.~Bandyopadhyay, J.~Silk, J.~E.~Taylor and T.~J.~Maccarone,
  Mon.\ Not.\ Roy.\ Astron.\ Soc.\ {\bf 392}, 1115 (2009);
  arXiv:0810.3674 [astro-ph].

  K.~S.~Cheng, D.~O.~Chernyshov and V.~A.~Dogiel,
  Astrophys.\ J.\  {\bf 645}, 1138 (2006)
  [arXiv:astro-ph/0603659].

 A.~Calvez and A.~Kusenko,
  arXiv:1003.0045 [astro-ph.HE].

\bibitem{higdon}

  J.~C.~Higdon, R.~E.~Lingenfelter and R.~E.~Rothschild,
  Astrophys.\ J.\  {\bf 698}, 350 (2009)
  [arXiv:0711.3008 [astro-ph]];
  Phys.\ Rev.\ Lett.\  {\bf 103}, 031301 (2009)
  [arXiv:0904.1025 [astro-ph.HE]].


\bibitem{asym}
  G.~Weidenspointner {\it et al.},
  Nature {\bf 451}, 159 (2008).

\bibitem{bouchet2}
  L.~Bouchet, J.~P.~Roques and E.~Jourdain,
  arXiv:1007.4753 [astro-ph.HE].


\bibitem{mevdm}

   C.~Boehm, T.~A.~Ensslin and J.~Silk,
  J.\ Phys.\ G {\bf 30}, 279 (2004)
  [arXiv:astro-ph/0208458].


  C.~Boehm and P.~Fayet,
  Nucl.\ Phys.\  B {\bf 683}, 219 (2004)
  [arXiv:hep-ph/0305261].




  C.~Boehm, D.~Hooper, J.~Silk, M.~Casse and J.~Paul,
  Phys.\ Rev.\ Lett.\  {\bf 92}, 101301 (2004)
  [arXiv:astro-ph/0309686].


 C.~Boehm, P.~Fayet and J.~Silk,
  Phys.\ Rev.\  D {\bf 69}, 101302 (2004)
  [arXiv:hep-ph/0311143].

 D.~Hooper, F.~Ferrer, C.~Boehm, J.~Silk, J.~Paul, N.~W.~Evans and M.~Casse,
  Phys.\ Rev.\ Lett.\  {\bf 93} (2004) 161302
  [arXiv:astro-ph/0311150].


 C.~Boehm and Y.~Ascasibar,
  Phys.\ Rev.\  D {\bf 70}, 115013 (2004)
  [arXiv:hep-ph/0408213].

 P.~Fayet,
  Phys.\ Rev.\  D {\bf 70}, 023514 (2004)
  [arXiv:hep-ph/0403226].

 P.~D.~Serpico and G.~G.~Raffelt,
  Phys.\ Rev.\  D {\bf 70}, 043526 (2004)
  [arXiv:astro-ph/0403417].


J.~F.~Beacom, N.~F.~Bell and G.~Bertone,
  Phys.\ Rev.\ Lett.\  {\bf 94}, 171301 (2005)
  [arXiv:astro-ph/0409403].


 K.~Ahn and E.~Komatsu,
  Phys.\ Rev.\  D {\bf 71}, 021303 (2005)
  [arXiv:astro-ph/0412630];
  Phys.\ Rev.\  D {\bf 72}, 061301 (2005)
  [arXiv:astro-ph/0506520].

 Y.~Rasera, R.~Teyssier, P.~Sizun, B.~Cordier, J.~Paul, M.~Casse and P.~Fayet,
  Phys.\ Rev.\  D {\bf 73}, 103518 (2006)
  [arXiv:astro-ph/0507707].


 J.~F.~Gunion, D.~Hooper and B.~McElrath,
  Phys.\ Rev.\  D {\bf 73}, 015011 (2006)
  [arXiv:hep-ph/0509024].


 P.~Sizun, M.~Casse and S.~Schanne,
  Phys.\ Rev.\  D {\bf 74}, 063514 (2006)
  [arXiv:astro-ph/0607374].

 C.~Jacoby and S.~Nussinov,
  JHEP {\bf 0705}, 017 (2007)
  [arXiv:hep-ph/0703014].

 Y.~Kahn, M.~Schmitt and T.~M.~P.~Tait,
  Phys.\ Rev.\  D {\bf 78}, 115002 (2008)
  [arXiv:0712.0007 [hep-ph]].


\bibitem{Yuksel}
  J.~F.~Beacom and H.~Yuksel,
  Phys.\ Rev.\ Lett.\  {\bf 97}, 071102 (2006)
  [arXiv:astro-ph/0512411].

\bibitem{chern}
  D.~O.~Chernyshov, K.~S.~Cheng, V.~A.~Dogiel, C.~M.~Ko and W.~H.~Ip,
  arXiv:0912.0889 [astro-ph.GA].



\bibitem{twist}
  F.~Chen, J.~M.~Cline and A.~R.~Frey,
  Phys.\ Rev.\  D {\bf 79}, 063530 (2009)
  [arXiv:0901.4327 [hep-ph]].

\bibitem{pamela}
  O.~Adriani {\it et al.}  [PAMELA Collaboration],
  Nature {\bf 458}, 607 (2009)
  [arXiv:0810.4995 [astro-ph]].


\bibitem{atic}
  J.~Chang {\it et al.},
  Nature {\bf 456} (2008) 362.



\bibitem{Finkbeiner-metastable}
  D.~P.~Finkbeiner, T.~R.~Slatyer, N.~Weiner and I.~Yavin,
  JCAP {\bf 0909}, 037 (2009)
  [arXiv:0903.1037 [hep-ph]].


\bibitem{us}
  F.~Chen, J.~M.~Cline, A.~Fradette, A.~R.~Frey and C.~Rabideau,
  arXiv:0911.2222 [hep-ph].


\bibitem{xdm}
  D.~P.~Finkbeiner and N.~Weiner,
  Phys.\ Rev.\  D {\bf 76}, 083519 (2007)
  [arXiv:astro-ph/0702587].


\bibitem{PR}
  M.~Pospelov and A.~Ritz,
  Phys.\ Lett.\  B {\bf 651}, 208 (2007)
  [arXiv:hep-ph/0703128].


\bibitem{nima}
  N.~Arkani-Hamed, D.~P.~Finkbeiner, T.~R.~Slatyer and N.~Weiner,
  Phys.\ Rev.\  D {\bf 79}, 015014 (2009)
  [arXiv:0810.0713 [hep-ph]].

\bibitem{harnik}
  P.~W.~Graham, R.~Harnik, S.~Rajendran and P.~Saraswat,
  arXiv:1004.0937 [hep-ph].

  R.~Essig, J.~Kaplan, P.~Schuster and N.~Toro,
  arXiv:1004.0691 [hep-ph].

\bibitem{dama}
  R.~Bernabei {\it et al.}  [DAMA Collaboration],
  Eur.\ Phys.\ J.\  C {\bf 56}, 333 (2008)
  [arXiv:0804.2741 [astro-ph]];
  Eur.\ Phys.\ J.\  C {\bf 67}, 39 (2010)
  [arXiv:1002.1028 [astro-ph.GA]].

\bibitem{cogent}
  C.~E.~Aalseth {\it et al.}  [CoGeNT collaboration],
  arXiv:1002.4703 [astro-ph.CO].


\bibitem{idm}


 D.~Tucker-Smith and N.~Weiner,
  Phys.\ Rev.\  D {\bf 64}, 043502 (2001)
  [arXiv:hep-ph/0101138].

 D.~Tucker-Smith and N.~Weiner,
  Phys.\ Rev.\  D {\bf 72}, 063509 (2005)
  [arXiv:hep-ph/0402065].

  S.~Chang, G.~D.~Kribs, D.~Tucker-Smith and N.~Weiner,
  Phys.\ Rev.\  D {\bf 79}, 043513 (2009)
  [arXiv:0807.2250 [hep-ph]].


\bibitem{Batell-multicomponent}
  B.~Batell, M.~Pospelov and A.~Ritz,
  Phys.\ Rev.\  D {\bf 79}, 115019 (2009)
  [arXiv:0903.3396 [hep-ph]].

\bibitem{lightdm}

 A.~L.~Fitzpatrick, D.~Hooper and K.~M.~Zurek,
  Phys.\ Rev.\  D {\bf 81}, 115005 (2010)
  [arXiv:1003.0014 [hep-ph]].

  S.~Andreas, C.~Arina, T.~Hambye, F.~S.~Ling and M.~H.~G.~Tytgat,
  arXiv:1003.2595 [hep-ph].

 S.~Chang, J.~Liu, A.~Pierce, N.~Weiner and I.~Yavin,
  arXiv:1004.0697 [hep-ph].

R.~Foot,
  arXiv:1004.1424 [hep-ph].


 K.~J.~Bae, H.~D.~Kim and S.~Shin,
  arXiv:1005.5131 [hep-ph].

Y.~Mambrini,
  arXiv:1006.3318 [hep-ph].

\bibitem{hooper}
  D.~Hooper, J.~I.~Collar, J.~Hall and D.~McKinsey,
  arXiv:1007.1005 [hep-ph].

\bibitem{decays}
 D.~Hooper and L.~T.~Wang,
  Phys.\ Rev.\  D {\bf 70}, 063506 (2004)
  [arXiv:hep-ph/0402220].
  C.~Picciotto and M.~Pospelov,
  Phys.\ Lett.\  B {\bf 605}, 15 (2005)
  [arXiv:hep-ph/0402178].

\bibitem{BPR}
  B.~Batell, M.~Pospelov and A.~Ritz,
  Phys.\ Rev.\  D {\bf 80}, 095024 (2009)
  [arXiv:0906.5614 [hep-ph]].


\bibitem{Navarro}
  J.~F.~Navarro {\it et al.},
	Mon.\ Not.\ Roy.\ Astron.\ Soc.\ {\bf 402}, 21 (2010)
  arXiv:0810.1522 [astro-ph].



\bibitem{bdm}
  E.~Romano-Diaz, I.~Shlosman, Y.~Hoffman and C.~Heller,
  arXiv:0808.0195 [astro-ph].

  M.~G.~Abadi, J.~F.~Navarro, M.~Fardal, A.~Babul and M.~Steinmetz,
  arXiv:0902.2477 [astro-ph.GA].

 S.~E.~Pedrosa, P.~B.~Tissera and C.~Scannapieco,
  arXiv:0910.4380 [astro-ph.CO].


\bibitem{tissera}
  P.~B.~Tissera, S.~D.~M.~White, S.~Pedrosa and C.~Scannapieco,
	Mon.\ Not.\ Roy.\ Astron.\ Soc.\ {\bf 406}, 922 (2010); 
  arXiv:0911.2316 [astro-ph.CO].


\bibitem{pato}
  M.~Pato, O.~Agertz, G.~Bertone, B.~Moore and R.~Teyssier,
  arXiv:1006.1322 [astro-ph.HE].

\bibitem{salucci}
  P.~Salucci, F.~Nesti, G.~Gentile and C.~F.~Martins,
  ``The dark matter density at the Sun's location,''
  arXiv:1003.3101 [astro-ph.GA].



\bibitem{Abidin}
  Z.~Abidin, A.~Afanasev and C.~E.~Carlson,
  arXiv:1006.5444 [hep-ph].

\bibitem{Prantzos}
  N.~Prantzos,
  Astron.\ Astrophys.\  {\bf 449}, 869 (2006)
  [arXiv:astro-ph/0511190];
  New Astron.\ Rev.\  {\bf 52}, 457 (2008)
  [arXiv:0809.2491 [astro-ph]].


\bibitem{Jean}

 P.~Jean, W.~Gillard, A.~Marcowith and K.~Ferriere,
  arXiv:0909.4022 [astro-ph.HE].



\bibitem{nonabelian}
  F.~Chen, J.~M.~Cline and A.~R.~Frey,
  Phys.\ Rev.\  D {\bf 80}, 083516 (2009)
  [arXiv:0907.4746 [hep-ph]].



 \bibitem{sommerfeld}
A. Sommerfeld, ``\"Uber die Beugung und Bremsung der Elektronen'', Ann. Phys. 403, 257 (1931).

J.~Hisano, S.~Matsumoto and M.~M.~Nojiri,
  Phys.\ Rev.\ Lett.\  {92} (2004) 031303
  [arXiv: hep-ph/0307216].

J.~Hisano, S.~Matsumoto, M.~M.~Nojiri and O.~Saito,
  Phys.\ Rev.\  D {71} (2005) 063528
  [arXiv: hep-ph/0412403].

M.~Cirelli, A.~Strumia, M.~Tamburini,
  Nucl.\ Phys.\  B {\bf 787} (2007) 152
  [arXiv:0706.4071 [hep-ph]].

 M.~Lattanzi and J.~I.~Silk,
  arXiv:0812.0360 [astro-ph].

See also previous work in 
  K.~Belotsky, D.~Fargion, M.~Khlopov and R.~V.~Konoplich,
  Phys.\ Atom.\ Nucl.\  {71} (2008) 147
  [arXiv:hep-ph/0411093] and references therein. 
 




\bibitem{feng}
  J.~L.~Feng, M.~Kaplinghat and H.~B.~Yu,
  arXiv:1005.4678 [hep-ph].


\bibitem{slatyer-com}
T.\ Slatyer, private communication


\bibitem{semiann}
  F.~D'Eramo and J.~Thaler,
  JHEP {\bf 1006}, 109 (2010)
  [arXiv:1003.5912 [hep-ph]].


\bibitem{KT} Kolb and Turner, The Early Universe, Addison-Wesley
(1988).

\bibitem{klb}
  F.~J.~Kerr and D.~Lynden-Bell,
  Mon.\ Not.\ Roy.\ Astron.\ Soc.\  {\bf 221}, 1023 (1986).


\bibitem{CC}
  M.~Cirelli and J.~M.~Cline,
  arXiv:1005.1779 [hep-ph].


\bibitem{Churazov}
  E.~Churazov, R.~Sunyaev, S.~Sazonov, M.~Revnivtsev and D.~Varshalovich,
  Mon.\ Not.\ Roy.\ Astron.\ Soc.\  {\bf 357}, 1377 (2005)
  [arXiv:astro-ph/0411351].


\bibitem{spectral}
 P.~Jean {\it et al.},
  Astron.\ Astrophys.\  {\bf 445}, 579 (2006)
  [arXiv:astro-ph/0509298].

\bibitem{guessoum}
  N.~Guessoum, P.~Jean and W.~Gillard,
  arXiv:astro-ph/0504186.

\bibitem{Ferriere}
  K.~Ferriere, W.~Gillard and P.~Jean,
  Astron.\ Astrophys.\  {\bf 467}, 611 (2007)
  [arXiv:astro-ph/0702532].

\bibitem{sawada}
  T.~Sawada, T.~Hasegawa, T.~Handa and R.~J.~Cohen,
  Mon.\ Not.\ Roy.\ Astron.\ Soc.\  {\bf 349}, 1167 (2004)
  [arXiv:astro-ph/0401286].

\bibitem{burton}
   H.S.\ Liszt, W.B.\ Burton, Astrophys.\ J.\  {\bf 236}, 779 (1980)

\bibitem{Skinner}
  G.~K.~Skinner,
  arXiv:1009.2098 [astro-ph.GA].

\bibitem{CL}
 J.~M.~Cordes and T.~J.~W.~Lazio,
  arXiv:astro-ph/0207156.

\bibitem{vicky}
V.\ Kaspi, private communication

\bibitem{teegarden}
  B.~J.~Teegarden and K.~Watanabe,
  Astrophys.\ J.\  {\bf 646}, 965 (2006)
  [arXiv:astro-ph/0604277].

\bibitem{yuksel-kistler}
  H.~Yuksel and M.~D.~Kistler,
  Phys.\ Rev.\  D {\bf 78}, 023502 (2008)
  [arXiv:0711.2906 [astro-ph]].

\bibitem{chandra}
  M.~P.~Muno {\it et al.},
  Astrophys.\ J.\  {\bf 613}, 326 (2004)
  [arXiv:astro-ph/0402087].

\bibitem{baum}
  M.~Baumgart, C.~Cheung, J.~T.~Ruderman, L.~T.~Wang and I.~Yavin,
  JHEP {\bf 0904}, 014 (2009)
  [arXiv:0901.0283 [hep-ph]].


\bibitem{arina}
  C.~Arina and M.~H.~G.~Tytgat,
  arXiv:1007.2765 [astro-ph.CO].

\bibitem{lavalle}
  J.~Lavalle,
  arXiv:1007.5253 [astro-ph.HE].

\bibitem{cmb}
 S.~Galli, F.~Iocco, G.~Bertone and A.~Melchiorri,
  Phys.\ Rev.\  D {\bf 80}, 023505 (2009)
  [arXiv:0905.0003 [astro-ph.CO]].


  T.~R.~Slatyer, N.~Padmanabhan and D.~P.~Finkbeiner,
  Phys.\ Rev.\  D {\bf 80}, 043526 (2009)
  [arXiv:0906.1197 [astro-ph.CO]].

\bibitem{CMB}
  M.~Cirelli, F.~Iocco and P.~Panci,
  JCAP {\bf 0910}, 009 (2009)
  [arXiv:0907.0719 [astro-ph.CO]].


\bibitem{posp-prad}
  M.~Pospelov and J.~Pradler,
  arXiv:1006.4172 [hep-ph].

\bibitem{feyncalc}
  R.~Mertig, M.~Bohm and A.~Denner,
  Comput.\ Phys.\ Commun.\  {\bf 64}, 345 (1991).



\bibitem{toro}
  J.~D.~Bjorken, R.~Essig, P.~Schuster and N.~Toro,
  Phys.\ Rev.\  D {\bf 80}, 075018 (2009)
  [arXiv:0906.0580 [hep-ph]].

R.~Essig, P.~Schuster, N.~Toro and B.~Wojtsekhowski,
  arXiv:1001.2557 [hep-ph].


\bibitem{PDG}
  K. Nakamura {\it et al.}, Journal of Physics {\bf G 37}, 075021 (2010) 


















\end{thebibliography}
\end{document}